\documentclass[twocolumn,tighten,times]{aastex631}
\usepackage{xcolor}
\usepackage{apjfonts}

\newcommand{\cajwl}{$Ca_{\rm JWL}$}
\newcommand{\chjwl}{$ch_{\rm JWL}$}

\newcommand{\cnjwl}{$cn_{\rm JWL}$}

\newcommand{\nrgb}{$n$(FG):$n$(SG)}

\newcommand{\pchjwl}{$\parallel$\,$ch_{\rm JWL}$}

\newcommand{\pcnjwl}{$\parallel$\,$cn_{\rm JWL}$}

\newcommand{\dchcn}{$\parallel$\,$ch_{\rm JWL} - \parallel$\,$cn_{\rm JWL}$}
\newcommand{\str}{Str\"omgren}
\newcommand{\vbump}{$V_{\rm bump}$}

\newcommand{\vvhb}{$V - V_{\rm HB}$}
\newcommand{\vvhbmag}{$-$2.0 mag $\leq$ $V - V_{\rm HB}$ $\leq$ 1.5 mag}
\newcommand{\cnwave}{$\lambda$3883}
\newcommand{\chwave}{$\lambda$4250}

\newcommand{\dy}{$\Delta Y$}
\newcommand{\hkjwl}{$hk_{\rm JWL}$}

\newcommand{\teff}{$T_{\rm eff}$}

\newcommand{\fehhk}{[Fe/H]$_{hk}$}
\newcommand{\cfech}{[C/Fe]$_{ch}$}
\newcommand{\nfecn}{[N/Fe]$_{cn}$}

\newcommand{\cfe}{[C/Fe]}
\newcommand{\nfe}{[N/Fe]}
\newcommand{\feh}{[Fe/H]}
\newcommand{\scfe}{$\sigma$[C/Fe]}
\newcommand{\scnfe}{$\sigma$[C,N/Fe]}
\newcommand{\scpnfe}{$\sigma$[C+N/Fe]}
\newcommand{\sfecnfe}{$\sigma$[Fe/H, C,N/Fe, C+N/Fe]}
\newcommand{\snfe}{$\sigma$[N/Fe]}
\newcommand{\dsfeh}{$\Delta\sigma$[Fe/H]}
\newcommand{\dscfe}{$\Delta\sigma$[C/Fe]}
\newcommand{\dsnfe}{$\Delta\sigma$[N/Fe]}
\newcommand{\dscnfe}{$\Delta\sigma$[C,N/Fe]}
\newcommand{\dscpnfe}{$\Delta\sigma$[C+N/Fe]}
\newcommand{\dsfecnfe}{$\Delta\sigma$[Fe/H, C,N/Fe, C+N/Fe]}
\newcommand{\sfeh}{$\sigma$[Fe/H]}

\newcommand{\dfehfg}{$\delta$[Fe/H]$_{\rm FG}$}
\newcommand{\dfehsg}{$\delta$[Fe/H]$_{\rm SG}$}

\newcommand{\ebv}{$E(B-V)$}
\newcommand{\eby}{$E(b-y)$}

\newcommand{\cfemv}{$d\mathrm{[C/Fe]}/M_V$}

\newcommand{\dmz}{$(m-M)_0$}
\newcommand{\dm}{$(m-M)$}
\newcommand{\dmv}{$(m-M)_V$}
\newcommand{\by}{$(b-y)$}
\newcommand{\fgfrac}{$N_{\rm FG}/N_{\rm tot}$}
\newcommand{\rgbw}{$\Delta$W$_{\rm CF336W,F438W,F814W}$}
\newcommand{\rgbwo}{W$_{\rm CF336W,F438W,F814W}$}
\newcommand{\logmass}{$\log M/M_{\rm \odot}$}

\newcommand{\linemake}{\texttt{Linemake}}
\newcommand{\moogscat}{\texttt{MOOG\_SCAT}}
\newcommand{\atlas}{\texttt{ATLAS12}}

\shortauthors{Lee et al.}
\shorttitle{NGC\,2257}

\begin{document}

\title{Multiple Populations of the Large Magellanic Cloud Globular Cluster NGC 2257:  No Major Environmental Effect on the Formation of Multiple Populations of the Old Globular Clusters in  Large Magellanic Cloud}

\author[0000-0002-2122-3030]{Jae-Woo Lee}
\affiliation{Department of Physics and Astronomy, Sejong University, 209 Neungdong-ro, Gwangjin-Gu, Seoul 05006, Republic of Korea,
jaewoolee@sejong.ac.kr, jaewoolee@sejong.edu}

\author{Tae-Hyeong Kim}
\affiliation{Department of Physics and Astronomy, Sejong University, 209 Neungdong-ro, Gwangjin-Gu, Seoul 05006, Republic of Korea}

\author[0000-0001-7033-4522]{Hak-Sub Kim}
\affiliation{Korea AseroSpace Administration, 537, Haeansaneop-ro, Sanam-myeon, Sacheon-si, Gyeongsangnam-do, 52535, Republic of Korea}

\author[0000-0001-9515-3584]{Hyun-Il Sung}
\affiliation{Korea Astronomy and Space Science Institute, 776 Daedeokdae-ro, Yuseong-gu, Daejeon 34055, Republic of Korea}

\author[0000-0003-4770-688X]{Hwihyun Kim}
\affiliation{International Gemini Observatory/NSF NOIRLab,\footnote{Supported by the international Gemini Observatory, a program of NSF NOIRLab, which is managed by the Association of Universities for Research in Astronomy (AURA) under a cooperative agreement with the U.S. National Science Foundation, on behalf of the Gemini partnership of Argentina, Brazil, Canada, Chile, the Republic of Korea, and the United States of America.} 950 N Cherry Ave, Tucson, AZ 85719, USA}

\author{Francesco Di Mille}
\affiliation{Las Campanas Observatory – Carnegie Institution for Science, Colina el Pino, Casilla 601, La Serena, Chile}

\begin{abstract}
How the environment of the host galaxy affects the formation of multiple populations (MPs) in globular clusters (GCs) is one of the outstanding questions in the near-field cosmology. To understand the true nature of the old GC MPs in the Large Magellanic Cloud (LMC), we study the Ca--CN--CH photometry of the old metal-poor LMC GC NGC\,2257. We find the predominantly FG-dominated populational number ratio of  \nrgb\ = 61:39($\pm$4), where the FG and SG denote the first and second generations. Both the FG and SG have similar cumulative radial distributions, consistent with the idea that NGC\,2257 is dynamically old. We obtain \fehhk\ = $-$1.78$\pm$0.00 dex($\sigma$=0.05 dex) and our metallicity is $\sim$0.2 dex larger than that from the high-resolution spectroscopy by other, due to their significantly lower temperatures by $\sim$ $-$200 K. The NGC\,2257 FG shows a somewhat larger metallicity variation than the SG, the first detection of such phenomenon in an old LMC GC, similar to Galactic GCs with MPs, strongly suggesting that it is a general characteristic of GCs with MPs. Interestingly, the NGC\,2257 SG does not show a helium enhancement compared to the FG. Our results for the Galactic normal GCs exhibit that the degree of carbon and nitrogen variations are tightly correlated with the GC mass, while NGC\,2257 exhibits slightly smaller variations for its mass. We show that old LMC GCs follow the same trends as the Galactic normal GCs in the \rgbw, \fgfrac, and \logmass\ domains. Our result indicates that the environment of the host galaxy did not play a major role in the formation and evolution of GC MPs.
\end{abstract}


\section{INTRODUCTION}
The Galactic globular clusters (GCs) exhibits multiple populations (MPs) showing elemental abundance patterns characterized by proton capture process at high temperatures, such as C--N, Na--O, and Mg--Al anticorrelations. The existence of the GC MPs is an ubiquitous but a mysterious nature \citep[e.g.,][]{carretta09, lee09, milone17}.  The typical Galactic normal GCs are composed of 30--40\% of the first generation (FG), whose lighter elemental abundance patterns are same as those of the Galactic halo field stars with similar metallicities, and 60--70\% of the second generation (SG), with abnormal elemental abundance patterns that experienced proton capture processes. The study of the GC MPs is a cornerstone of galactic archaeology, which will provide critical insights into the formation of the Galactic halo, offering clues to the early assembly and chemical evolution of our Galaxy, for example.  Unfortunately, in spite of tremendous effort directed at understanding the formation of GC MPs both in the theoretical and observational aspects, no satisfactory explanation yet exists so far \citep[e.g.,][]{dercole08, bastian18, gratton19}

The recent studies showed that GCs not only in our Galaxy but also in nearby galaxies exhibit MPs \citep[e.g.,][]{larsen14, hollyhead17, niederhofer17a, niederhofer17b, milone20}. In particular, GCs in Magellanic Clouds (MCs) show that the FG is the major component taking up to 60--70\%, somewhat larger than the Galactic counterparts with the similar metallicity, $\sim$30\% \citep{milone20}, which may suggest that the host galaxy environments are most likely affect the formation and evolution of GCs. In our Galaxy, the bifurcation of the age--metallicity relation between the in situ and ex situ GCs could a good example telling that the physical environments of the ex situ GCs' host galaxies must have been different from our Galaxy \citep[e.g.,][]{massari19, fobes20}. Another good example would be the quite different GC systems in the Large Magellanic Cloud (LMC) and the Small Magellanic Cloud (SMC): The number of clusters in the SMC is much fewer than the LMC is and there is no very old (age $\lesssim$ 11 Gyr) metal-poor (\feh\ $\lesssim$ $-$1.5 dex) GC in the SMC.

The LMC is the most massive satellite galaxy associated with the Milky Way. It is well-known that the stellar contents of the LMC do not fit to the scaled-down relations of those found in our Galaxy. For example, the LMC GCs exhibit the famous age-gap, no GCs with the age from $\sim$4 to $\sim$9 Gyrs \citep[e.g.,][]{geisler97, dacosta02}: The young clusters in the LMC are all compact and are located preferentially in the inner part, while the old clusters show diverse sizes and locations. The age of the oldest LMC GCs is comparable to that of the MW GCs \citep[e.g.,][]{johnson99, milone23}. However, regardless of ages, metallicities and locations, the most of the LMC GCs show disk-like kinematics, in sharp contrast to the Galactic GCs showing both the disk-like and halo-like kinematics \citep[e.g.,][]{bennet22}.  Nonetheless, studies of the LMC GCs would provide an unique  opportunity to understand how the mass of the host galaxy affects the physical properties of the old GC system.

The properties of GC MPs have been studied using a large sample size mostly by employing HST broad band photometry. For example, \citet{milone17} argued that the RGB color width of Galactic GCs is correlated with metallicity. Furthermore, they showed that the RGB width is significantly correlated with the mass and luminosity of GCs after removing the metallicity dependency. Later, \citet{lagioia19b} found the similar results for the MC GCs; the metallicity and mass dependency on the width of the intrinsic RGB pseudo-color. They also found that the MC GCs exhibit systematically narrower RGB widths than those of Galactic GCs with similar mass and metallicity, suggesting that MC GCs might have smaller internal light elemental abundance variations than the Galactic GCs. We note that the ages of MC GCs studied by \citet{lagioia19b} are significantly younger  ($\lesssim$ 5.4 Gyr) than those of Galactic GCs. More recently, \citet{vanaraj21} studied older MC GCs and they argued that the two old LMC GCs, NGC\,1786 \citep[12.9 Gyr,][]{milone23} and NGC\,1898 \citep[11.7 Gyr,][]{milone23}, follow the same trend without any offset values as the Galactic GCs, leading them to suggest that the host galaxy environment may play a minor role in the formation of GC MPs.

NGC\,2257 is the old metal-poor LMC GCs \citep[11.8 Gyr,][]{milone23}. Due to its favorable location on the sky (decl.\ $\sim -64^\circ$) for the LMC objects and a small interstellar reddening value, it has been frequently studied for decades. One of the early noteworthy study was done by \citet{walker89}, who found that it is a RR Lyrae (RRL) rich and its horizontal branch morphology is very similar to that of M3 \citep[see also][]{olszewski96}. He also argued that the metallicity of NGC\,2257 is \feh\ $\sim-$1.8 dex and it suffers a very small amount of interstellar reddening, \ebv\ $\sim$ 0.04 mag. Similarly, other previous studies indicate that the metallicity of NGC\,2257 is similar to or slightly metal-poor than M3. \citet{dirsch00} obtained \feh\ = $-1.63\pm$0.21 dex for NGC\,2257 using the \str\ photometry. Later, \cite{grocholski06} employed the low-resolution multi-object spectroscopy to obtain the infrared \ion{Ca}{2} triplets absorption strengths for 16 NGC\,2257 red-giant branch (RGB) stars and obtained \feh$_{\rm CG97}$ = $-1.59\pm0.02$ dex in the metallicity scale proposed by \citet{cg97}. A similar conclusion was drawn by \citet{song21}, \feh$_{\rm CG97}$ = $-1.59\pm0.02$ dex, using the low S/N single-order spectra.  Recently, \citet{sarajedini24} derived metallicities for old LMC GCs using the mean period of fundamental RRL stars and he obtained \feh\ of NGC\,2257 ranging from $-$1.63$\pm$0.014 to $-$1.94$\pm$0.04, depending on the adopted RRL dataset and metallicity--period relation. On the other hand, \citet{mucciarelli10} employed high- and intermediate-resolution spectroscopy, and they claimed \feh\ = $-$1.95$\pm$0.02 dex, significantly lower than previous measurements. We will argue that they have significantly lower effective temperatures by $\sim$ 200 K and their metallicity should be increased by $\sim$ 0.2 dex.

%%%%%%%%%%%%%%%%%%%%%%%%%%%%%%%%%%%%%%%%%%%%%%%%%%%%%%%%%%%%%%
\begin{figure*}
\epsscale{1.}
\figurenum{1}
\plotone{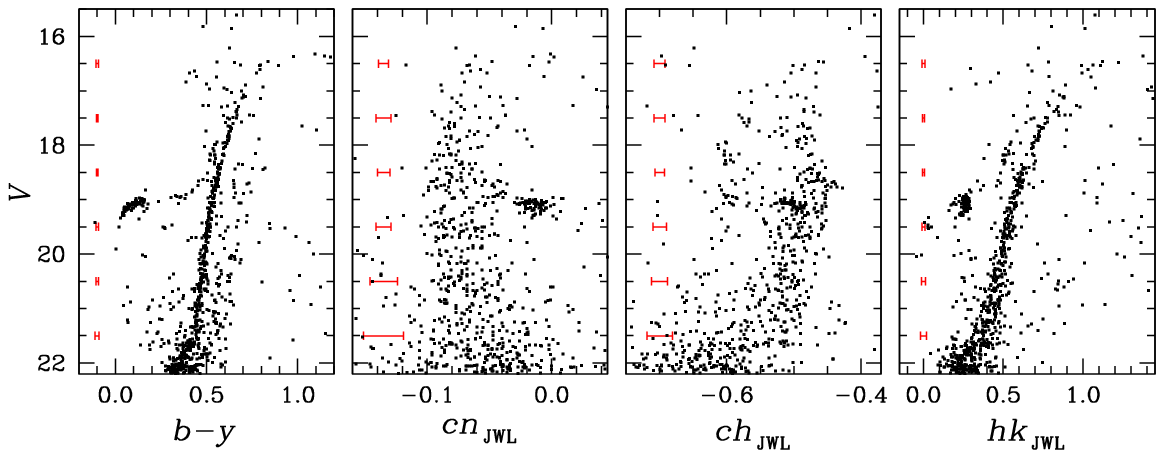}
\caption{CMDs of NGC\,2257 for the clean stars. The red error bars indicate the measurement uncertainties at a given $V$ magnitude.}\label{fig:cmd}
\end{figure*}
%%%%%%%%%%%%%%%%%%%%%%%%%%%%%%%%%%%%%%%%%%%%%%%%%%%%%%%%%%%%%%

The true MP nature of NGC\,2257 is somewhat ambiguous. The populational number ratio inferred from the oxygen and sodium abundances from \citet{mucciarelli10} may suggest that \nrgb\ = 2:4, a bona fide SG dominated GC similar to normal Galactic GCs, although their sample size is small. On the other hand, \citet{gilligan20} used the HST archival data for F336W, F606W, and F814W passbands to study the MPs in NGC\,2257. Albeit inadequacy of their datasets to study MPs in metal-poor GCs \citep[e.g., see][]{lee24}, they claimed to find the SG fraction of 30--40\% in the upper main-sequence (MS), which is in marginally agreement with our result as will be discussed later. Curiously, they found no evidence of MPs in the RGB and HB, where the presence of MPs is supposed to be more prominent than the upper MS, of NGC\,2257.

In this paper, we present a Ca--CN--CH photometry of NGC\,2257. In our previous studies, we elaborately demonstrated that our exquisite photometric system works excellently for the MPs in Galactic GCs. In particular, our system can simultaneously measure accurate \feh, \cfe, and \nfe, the key elements to understand the evolution of the low-mass stars and the MPs in GCs \citep[e.g.,][]{lee24}. Furthermore, with stringent population classification from our unique photometric system, we could be able to evaluate more meaningful physical properties of individual MPs. Here, we look into the carbon and nitrogen abundance variations and populational number ratios of NGC\,2257 and Galactic GCs, and speculate how the GC mass and host galaxies affect the formation of GC MPs.

\section{OBSERVATIONS AND DATA REDUCTION}\label{s:reduction}

Observations for NGC\,2257 were made on 2021 December 8 and 2022 January 6 using the Gemini-South 8.0 m telescope and the Gemini Multi-Object Spectrograph (GMOS) with our own filter system. The GMOS is a spectro-imager offering a spectral range from 0.36 to 1.03 micrometers across a field of view (FOV) spanning 5.5 square arcminutes. Its spatial sampling is set at 0.0807\arcsec\ per pixel and the observations of our science frames were done with a 2$\times$2 binning. We used our own custom-designed filters, including  \str\ $y$, $b$, \cajwl, JWL39 and JWL43 \citep{lee15, lee17, lee19b}. It is unfortunate that the Hamamatsu CCDs used in the GMOS-South are optimized for red wavelengths providing a quantum efficiency (QE) of $\sim$ 90-96\% in the 5000--7600 \AA\ range, compared to a lower QE of 33-52\% in the 3400--4000 \AA\ range as detailed in the GMOS Components page.\footnote{https://www.gemini.edu/instrumentation/gmos/components\#GSHam}
In addition, the main mirror of the Gemini South is silver-coated to enhance the sensitivity of the mid-infrared instruments. Consequently, some of our filters fall within the lower QE range, which may impact the sensitivity of our measurements in those spectral regions. Due to this reason, we were not able to achieve good signal-to-noise ratios for our JWL34 \citep[e.g., see][]{lee21a} and we did not use it.

During the observing runs for NGC\,2257, the average seeing was 1.04$\pm$0.12 arcsec. The total integration times were 2535 s, 6870 s, 29133 s, 13800 s, and 9000 s for \str\ $y$, $b$, \cajwl, JWL39 and JWL43, respectively.

The raw data were reduced with \texttt{DRAGONS 3.1.0} \citep{dragons}, a Python-based data reduction pipeline provided by the Gemini Observatory. The software identified and subtracted bias frames from the raw data, and generated master flat-filed frames from a subset of twilight flat exposures taken around the observation dates. The master flat-field frames were then used to correct for pixel-to-pixel sensitivity variations.

Photometry for NGC\,2257 were performed using \texttt{DAOPHOTII}, \texttt{DAOGROW}, \texttt{ALLSTAR} and \texttt{ALLFRAME}, and the relevant packages \citep{pbs87, pbs94}. It was very unfortunate that the photometric data collected through the Gemini queue observations did not allow us to derive atmospheric extinctions for individual filters. Therefore, we calculated the atmospheric extinction coefficients for our individual filter passbands using the best fit atmospheric extinction curve for Paranal by \citet{patat11}. This may well cause small offsets in our photometric zero-points, since the observations of NGC\,2257 were made at rather large airmasses. As we will show below, this is not the case and our photometric zero-points are in excellent agreement with those of others. The photometric transformations were applied using the Galactic GCs with known our photometry taken on the same night as photometric standards.  Finally, the astrometric solutions for individual stars were calculated using the Gaia Data Release 3 \citep[DR3;][]{gaiadr3} and the \texttt{IRAF IMCOORS} package.

\section{Results}
\subsection{Photometric Indices and Color--Magnitude Diagrams}
Throughout this work, we will use our own photometric indices \citep[see also][]{lee19a, lee21a, lee22}, defined as
\begin{eqnarray}
{{hk}}_{\mathrm{JWL}} &=& ({\mathrm{Ca}}_{\mathrm{JWL}}-b)-(b-y),\\
{{cn}}_{\mathrm{JWL}} &=& {JWL}39-{\mathrm{Ca}}_{\mathrm{JWL}},\\
{{ch}}_{\mathrm{JWL}} &=& ({JWL}43-b)-(b-y).
\end{eqnarray}
The \hkjwl\ index is a good photometric measure of metallicity. The \cnjwl\ and \chjwl\ indices are measures of CN at \cnwave\ and CH at \chwave\ \AA, respectively  \citep[e.g., see][]{lee09, lee15, lee22, lee23a, lee21a}.

In Figure~\ref{fig:cmd}, we show color--magnitude diagrams (CMDs) of NGC\,2257 using our color indices. In the figure, we also show the relevant measurement uncertainties for good quality stars with clean neighborhoods using the separation index \citep{pbs03}. We note that the RGB widths in the \cnjwl\ and \chjwl\ indices are broader than our measurement uncertainties, in particular for stars with $V \lesssim$ 21 mag, in our CMDs. These strongly suggest that the variations in carbon and nitrogen abundances in the RGB stars are responsible for the broad \cnjwl\ and \chjwl\ RGB sequences given that the differential reddening effect across our science frame is negligible, as we will discuss below.

%%%%%%%%%%%%%%%%%%%%%%%%%%%%%%%%%%%%%%%%%%%%%%%%%%%%%%%%%%%%%%
\begin{figure}
\epsscale{1.15}
\figurenum{2}
\plotone{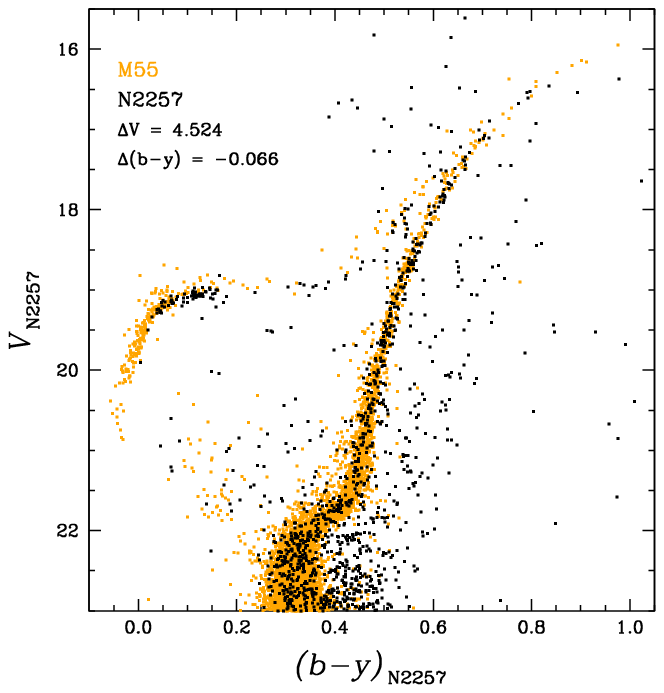}
\caption{Comparison of NGC\,2257 (black dots) and M55 (orange dots). The $V$ magnitude and \by\ color of individual M55 stars are increased by $\Delta V$ = 4.524 and $\Delta$\by\ = 0.066 mag.}\label{fig:wrtm55}
\end{figure}
%%%%%%%%%%%%%%%%%%%%%%%%%%%%%%%%%%%%%%%%%%%%%%%%%%%%%%%%%%%%%%

%%%%%%%%%%%%%%%%%%%%%%%%%%%%%%%%%%%%%%%%%%%%%%%%%%%%%%%%%%%%%%
\begin{figure*}
\epsscale{1.}
\figurenum{3}
\plotone{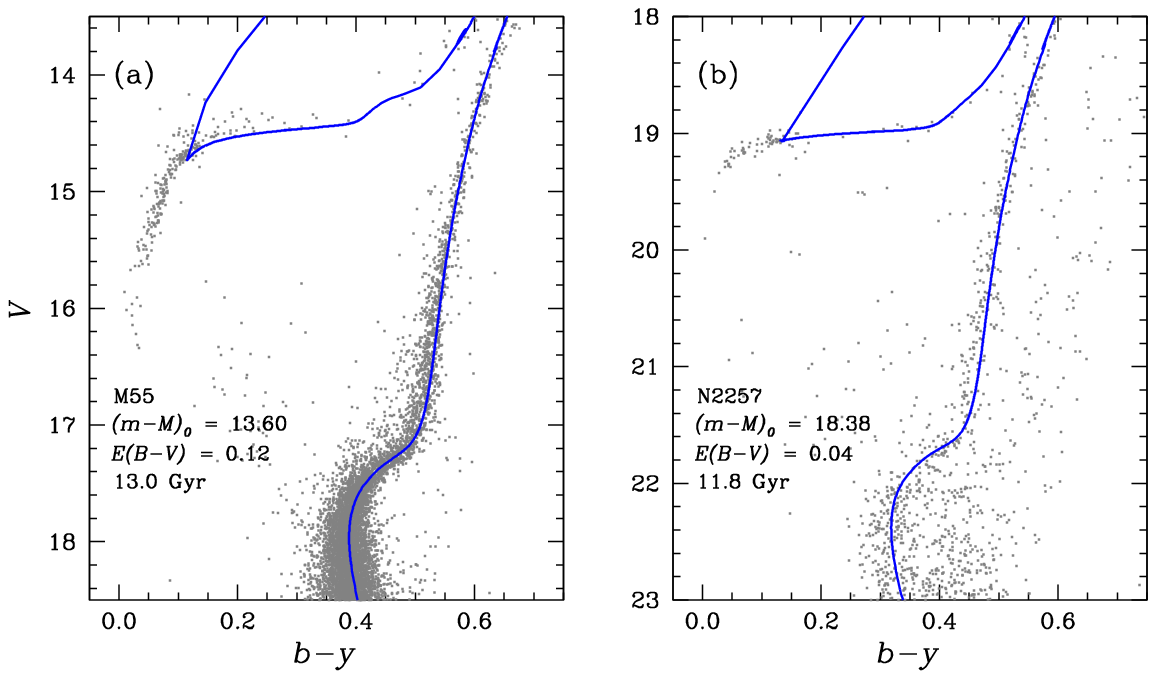}
\caption{(a) Comparison of M55 and the model isochrone for \feh\ = $-$1.8 dex and 13.0 Gyr. The distance modulus of \dm\ = 13.80 mag and the interstellar reddening value of \ebv\ = 0.12 mag are used. (b) Comparison of NGC\,2257 and the model isochrone for \feh\ = $-$1.8 dex and 11.8 Gyr. The distance modulus of \dmz\ = 18.38 mag and the interstellar reddening value of \ebv\ = 0.04 mag are used.}\label{fig:compiso}
\end{figure*}
%%%%%%%%%%%%%%%%%%%%%%%%%%%%%%%%%%%%%%%%%%%%%%%%%%%%%%%%%%%%%%

\subsection{Distance Modulus and Interstellar Reddening}
Due to the paucity of main-sequence turn-off (MSTO) stars and the contamination from the off-cluster stars in our NGC\,2257 CMDs, we learn that it is a bit difficult to define the accurate main sequence turn-off (MSTO) of NGC\,2257. Instead of comparing model isochrones to NGC\,2257, we attempted to derive relative distance modulus and interstellar reddening value of NGC\,2257 with respect to that of M55.\footnote{As we will show below, the metallicity of NGC\,2257 is similar to those of M55, \feh\ $\sim$ $-$1.8 dex \citep{carretta09, lee16, gontcharov23}.}

In Figure~\ref{fig:wrtm55}, we show \by\ CMDs of NGC\,2257 and M55 \citep{lee15, lee16}. For M55, we applied offset values of $\Delta V$ = 4.524 mag and $\Delta$\by\ = $-$0.066 mag to match its blue horizontal branch (BHB) and RGB to those of NGC\,2257. This relative reddening value is corresponding to $\Delta$\ebv\ = $-$0.080 mag between NGC\,2257 and M55  assuming \eby\ = 0.82 \ebv\ \citep{lee23b}, in the sense that M55 suffers from more severe interstellar reddening.

Recently, \citet{gontcharov23} studied M55 using 23 photometric passbands, finding \dmz\ = 13.60$\pm$0.01 mag, \dmv\ = 13.97$\pm$0.03 mag, and  \ebv\ = 0.118$\pm$0.004 mag \citep{gontcharov23}. Using their results for M55, we obtained \dmv\ = 18.494$\pm$0.033 mag and \ebv\ = 0.038$\pm$0.006 mag for NGC\,2257. Assuming $R_V$ = 3.1, we have \dmz\ = 18.376$\pm$0.019 mag. Our distance modulus and interstellar reddening for NGC\,2257 are in excellent agreement with those of \citet{milone23}, \dmz\ = 18.37 mag and  \ebv\ = 0.04 mag, who made use of the HST photometry to derive the cluster's parameters. This is a strong indication that our photometric zero-points are highly reliable.

In Figure~\ref{fig:wrtm55}, we would like to point out that the NGC\,2257 BHB stars occupy a very small region in the \by\ color with respect to M55, reminiscent of the bona fide second parameter pair of M30 and M92 \citep[see Figure~1 of][]{lee24}. This may reflect a very small internal helium dispersion among different popoulations as we will discuss below.

In Figure~\ref{fig:compiso}, we show comparisons of M55 and NGC\,2257 with relevant model isochrones from a Bag of Stellar Tracks and Isochrones \citep[BaSTI;][]{basti21} with distance moduli and interstellar reddening values that we discussed above. Our relative distance modulus and interstellar reddening for NGC\,2257 provide an excellent agreement with the model isochrone, strongly suggesting that our results are correct.

%%%%%%%%%%%%%%%%%%%%%%%%%%%%%%%%%%%%%%%%%%%%%%%%%%%%%%%%%%%%%%
\begin{figure*}
\epsscale{1.}
\figurenum{4}
\plotone{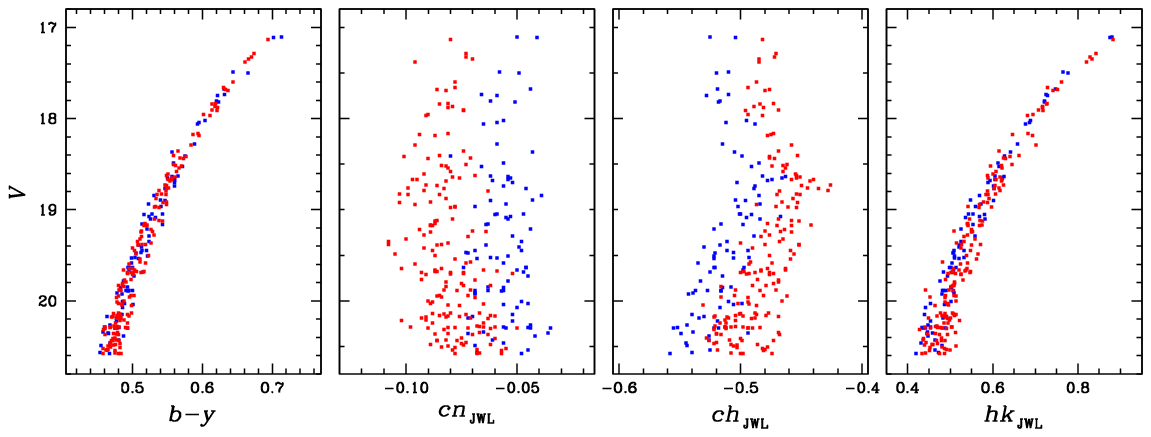}
\caption{CMDs of NGC\,2257 with \vvhbmag. The blue and red dots denote the FG and SG RGB stars returned from our EM estimator discussed below.}\label{fig:rgbpop}
\end{figure*}
%%%%%%%%%%%%%%%%%%%%%%%%%%%%%%%%%%%%%%%%%%%%%%%%%%%%%%%%%%%%%%

%%%%%%%%%%%%%%%%%%%%%%%%%%%%%%%%%%%%%%%%%%%%%%%%%%%%%%%%%%%%%%
\begin{figure}
\epsscale{1.2}
\figurenum{5}
\plotone{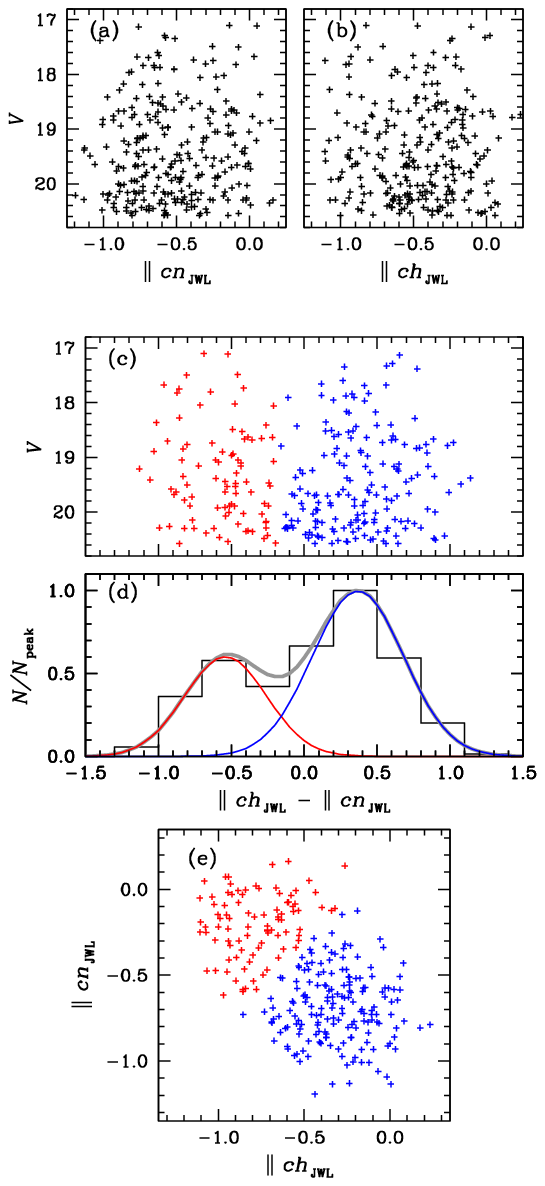}
\caption{(a) The \pchjwl\ vs. $V$ CMD of NGC\,2257 RGB stars.  (b) Same as (a) but for the  \pchjwl. (c) The (\dchcn) distribution of the NGC\,2257 RGB stars. The blue and red plus signs denote the FG and SG RGB stars in NGC\,2257 returned from our EM estimator.
(d) The histogram for the (\dchcn) distribution of the NGC\,2257 RGB stars. The blue and the red lines denote the FG and SG distributions returned from our EM estimator, while the gray line is for the all stars.
(e) Plot of the \pchjwl\ vs.\ \pcnjwl\ of NGC\,2257 RGB stars, showing a \pchjwl--\pcnjwl\ anticorrelation, a surrogate of the C--N anticorrelation.
}\label{fig:em}
\end{figure}
%%%%%%%%%%%%%%%%%%%%%%%%%%%%%%%%%%%%%%%%%%%%%%%%%%%%%%%%%%%%%%

\subsection{Populational Tagging for RGB Stars}
In order to perform populational tagging, we select all RGB stars with \vvhbmag.
We set the faint limit of \vvhb\ = 1.5 mag (i.e., $V \sim$ 20.5 mag), since our photometric measurement uncertainties becomes large for stars fainter than $V \sim$ 21.0 mag, in particular in the \cnjwl\ index as shown in Figure~\ref{fig:cmd}, due to the red sensitive nature of the Gemini-South telescope and the GMOS combination. Inclusion of stars fainter than \vvhb\ $\gtrsim$ 1.5 mag will bring large uncertainties in our elemental abundance derivations discussed below. On the other hand, we set the bright limit of \vvhb\ = $-$2.0 mag (i.e., $V \sim$ 17.0 mag) for two reasons. First, the number of stars near the RGB tip is very small and, therefore, these stars do not significantly influence the physical parameters that we will derive below in this work, such as the populational number ratio, the mean metallicity, the cumulative radial distributions, the RGB bumps and relative helium abundances, and etc. Second, in our previous studies for the Galactic GCs, we learned that the Kurucz stellar atmospheres \citep{kurucz11} near the RGB tip may not be accurate enough to properly reproduce some of our observed color indices with great satisfaction. At the same time, inappropriate treatment of the Rayleigh scattering from the neutral hydrogen (RSNH) and the carbon isotopic ratios can affect some of our synthetic color indices, in particular for the bright RGB stars in the metal-poor domain. Therefore, we try to exclude the brightest RGB stars in our analysis.

For our current work, the Gaia proper-motion study is not useful, since it is far from complete in our NGC\,2257 science field. Instead, we relied on our multi-color photometry for selecting member RGB stars, following the similar method that we employed previously for Galactic GCs \citep[e.g.,][]{lee15,lee17}. First, we  selected the RGB candidates around the isochrone in our $(b-y)$ CMD in Figure~\ref{fig:compiso}. Next, we excluded outliers in our \hkjwl, \cnjwl, and \chjwl\ CMDs. In our previous work \citep{lee15}, we discussed for M22 that our color indices, in particular the \hkjwl, are very effective to clear away the foreground and background field stars with different metallicities  \citep[see also,][]{lee20, lee23b}. Also, the \cnjwl\ and \chjwl\ absorption strengths have strong metallicity dependencies at given magnitudes and they are very useful to deselect off-cluster field stars with different metallicities \citep[e.g., see Figure~16 of][]{lee15}. The metallicity of NGC\,2257 is \feh\ $\sim$ $-$1.8 dex (see below) and it is significantly more metal-poor than the bulk of the LMC field stars, \feh\ $\sim-$0.6 dex \citep[e.g.,][]{cole00}. Therefore, the more metal-rich LMC field stars in our \hkjwl\ CMD can be easily distinguishable from the NGC\,2257 membership RGB stars.\footnote{See also Figure~\ref{fig:rgb} for a large \hkjwl\ difference between the RGB loci for \feh\ $\sim$ $-$1.8 and $-$0.7 dex.} Using Gaia photometry, we made a visual inspection of the stellar populations near our NGC\,2257 science field. We confirmed that our NGC\,2257 CMDs are contaminated mostly by the LMC field stars, which may assure validity of our cleaning procedure.

Due to internal mixing episodes accompanied by the CN cycle in the hydrogen shell burning region, the surface carbon and nitrogen abundances of the RGB stars brighter than the RGB bump (RGBB) are significantly changed from their initial abundances, which affect our \cnjwl\ and \chjwl\ indices as shown in Figure~\ref{fig:rgbpop}. Furthermore, the mixing efficiency sensitively depends on metallicity since the hydrogen shell will be hotter and wider with decreasing metallicity \citep[e.g., see][and references therein and see below]{lee23b}. Therefore, it is difficult to classify individual populations for the bright RGB stars from their observed color indices. To remove this luminosity effect, we make use of the normalized color indices \pcnjwl\ and \pchjwl, which are defined as
\begin{eqnarray}
\parallel cn_{\rm JWL} &\equiv& \frac{cn_{\rm JWL} - cn_{\rm JWL,red}}
{cn_{\rm JWL,red}-cn_{\rm JWL,blue}},\\
\parallel ch_{\rm JWL} &\equiv& \frac{ch_{\rm JWL} - ch_{\rm JWL,red}}
{ch_{\rm JWL,red}-ch_{\rm JWL,blue}},
\end{eqnarray}
where the subscripts denote the fiducials of the blue (the 5th percentile) and the red (the 95th percentile) sequences of individual color indices at given magnitudes \citep[see also][]{milone17, lee21a}.

In Figure~\ref{fig:em} (a--b), we show plots of the \pcnjwl\ and \pchjwl\ versus $V$ for RGB stars with \vvhbmag, where any bimodal distributions are difficult to see. Following the same procedure in our previous studies \citep[e.g., see][]{lee20}, we attempt to tag different populations on the (\dchcn) distribution, which is equivalent to the ([C/Fe] $-$ [N/Fe]) distribution in the Galactic normal GCs. In Figure~\ref{fig:em} (c--d), we show the plot of the (\dchcn) versus $V$ and a (\dchcn) histogram, clearly showing double peaks. We applied the expectation maximization (EM) algorithm for the multiple-component Gaussian mixture distribution model to the observed (\dchcn) distribution. We calculated the probability of individual RGB stars for being the FG (i.e., large \cfe\ and small \nfe) and SG (i.e., small \cfe\ and large \nfe) populations in an iterative manner, where stars with $P($FG$|x_i)$ $\geq$ 0.5  from the EM estimator are denoted with the solid blue line which corresponds to  the FG population, while $P($SG$|x_i)$ $>$ 0.5 with the solid red line which corresponds to the SG population. Through this process, we obtained the RGB populational number ratio of \nrgb\ = 61:39 ($\pm$4), being NGC\,2257 a significantly FG dominated GC.
Lastly, we show a plot of the \pchjwl\ versus \pcnjwl\ in Figure~\ref{fig:em}(e). As we will show below, NGC\,2257 RGB stars show a carbon and nitrogen anticorrelation as for the Galactic normal GCs. Therefore, our \cnjwl\ should be a measure of the nitrogen abundances in the RGB stars, provided that nitrogen is less abundant than carbon in NGC\,2257. As a result, this \pchjwl\ versus \pcnjwl\ anticorrelation is equivalent to the \cfe\ versus \nfe\ anticorrelation seen in the Galactic normal GCs.

We note the results by \citet{gilligan20}, who used the HST archival data to study the MPs in NGC\,2257. They used F336W, F606W, and F814W images taken with Advanced Camera for Surveys and Wide Field Camera 3, which are not ideal passbands to study MPs in metal-poor GCs. As \citet{lee24} discussed, the CN-red system, which is critical to classify populations, rapidly vanishes in the metal-poor domains due to the double-metallic nature of the CN molecule. The histograms presented by \citet{gilligan20} did not show any distinguishable bimodal peaks but slightly skewed distributions. They claimed the SG fraction of 30--40\% in the upper main-sequence, which is in agreement with our result. However, they found no evidence of MPs in the RGB and HB of NGC\,2257, which may indicate the drawback of the broad-band photometry for the GC MP study.

%%%%%%%%%%%%%%%%%%%%%%%%%%%%%%%%%%%%%%%%%%%%%%%%%%%%%%%%%%%%%%
\begin{figure}
\epsscale{1.1}
\figurenum{6}
\plotone{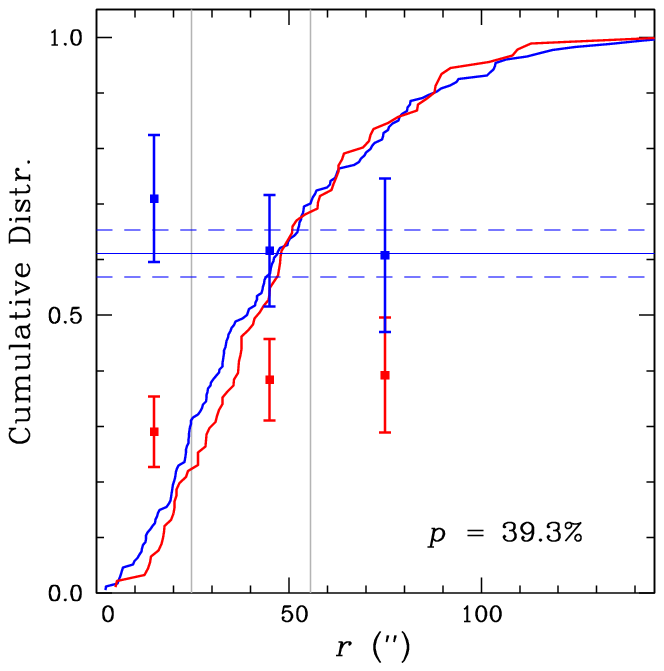}
\caption{The CRDs and populational number ratios of the FG (blue) and SG (red) in NGC\,2257. The horizontal solid line denotes the mean FG fraction, while the dashed lines the 1$\sigma$ error of the mean. The blue and red filled squares with error bars indicate the FG and SG populational number ratios with a radial bin size of 30\arcsec. The gray vertical lines denote the core and half-mass radii \citep{lanzoni19}.}\label{fig:crd}
\end{figure}
%%%%%%%%%%%%%%%%%%%%%%%%%%%%%%%%%%%%%%%%%%%%%%%%%%%%%%%%%%%%%%

\subsection{Cumulative Radial Distributions}
The observed cumulative radial distributions (CRDs) of individual GC MPs may depend on many physical properties at the time of GC formation and also provide critical information on the dynamical evolution of GCs.
Previous studies on the formation of the GC MPs suggested that the SG populations may form in the innermost part of the cluster in a more extended FG system \citep[e.g., see][]{dercole08, bekki19}, where the degree of the helium enhancement of the SG can be dependent on the external gas density \citep{calura19, lee21b}.
The initial structural difference between the FG and SG populations can be gradually erased out with time, due to the result of the preferential loss of the FG stars during the cluster's dynamical evolution \citep[e.g., see][]{vesperini21}.
It is also likely that the current location can be affected by the different evolutionary stellar masses of individual populations due to different degree of diffusion processes. Over the Hubble time, for example, the radial distributions of the later generations of stars with enhanced helium abundances can expand outward due to the smaller stellar masses resulted from their fast stellar evolution \citep[e.g., see][]{fare18}.

In this regards, we derive the CRDs of individual MPs in NGC\,2257, and we show our results in Figure~\ref{fig:crd}. It is very interesting to note that both populations appear to have similar CRDs. We performed the Kolmogorov--Smirnov (K--S) tests and we obtained the $p$ value of 0.393, suggesting that they are highly likely drawn from the same parent distribution.
We also calculate the populational number ratios for each population with a radial bin size of 30\arcsec, showing no substantial radial gradient in populational number ratios.

In their study, \citet{ferraro19} found that the blue straggler stars (BSSs), which are more massive than the bulk of normal stars in a given GC, are more centrally concentrated than the normal stars in NGC\,2257, due to the dynamical friction that brings massive BSSs toward the central part of the cluster, a strong indication of the dynamically old nature of NGC\,2257. Our similar CRDs between the two populations may support this. Presumably the asymmetric CRDs between the FG and SG at the formation epoch \citep{dercole08} can have enough dynamical time to become more relaxed system with the similar individual CRDs in NGC\,2257.

%%%%%%%%%%%%%%%%%%%%%%%%%%%%%%%%%%%%%%%%%%%%%%%%%%%%%%%%%%%%%%
\begin{figure}
\epsscale{1.1}
\figurenum{7}
\plotone{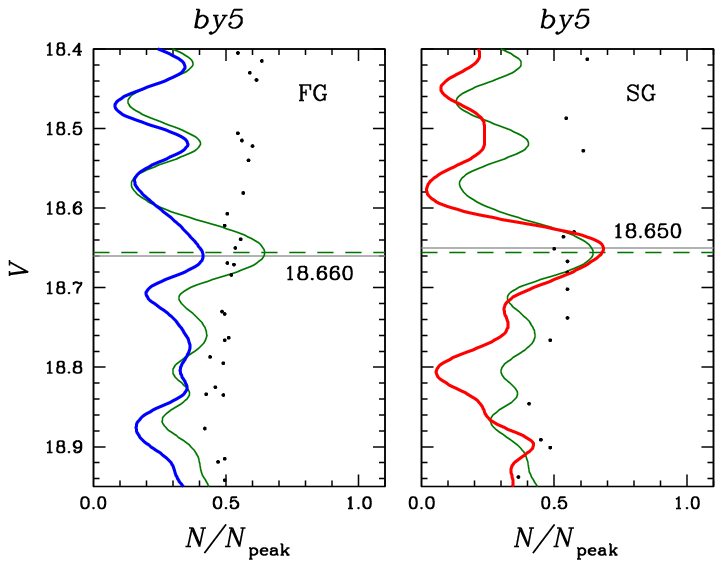}
\caption{Plots of the $by5$ vs.\ $V$ CMDs and differential LFs. The blue and red lines are LFs of the FG and SG, while the green lines are for the LF of the all stars. We also show the RGBB $V$ magnitudes of each populations with solid gray lines, while the dotted green lines denote the RGBB $V$ magnitude from all stars. Note that the $by5$ is defined to be $by5 = 5\times(b-y)-2.25$ for the clarity of the figure.}\label{fig:rgbb}
\end{figure}
%%%%%%%%%%%%%%%%%%%%%%%%%%%%%%%%%%%%%%%%%%%%%%%%%%%%%%%%%%%%%%

\subsection{Red-Giant Branch Bump}
During the course of low-mass star evolution, RGB stars experience a temporary drop in luminosity when the very thin H-burning shell crosses the discontinuity in the chemical composition and lowered mean molecular weight left by the deepest penetration of the convective envelope during the ascent of the RGB \citep[e.g.,][]{cassisi13}. In such evolutionary phases, RGB stars have to cross three times the same luminosity interval, leaving a distinctive feature, the so-called RGBB. It is well studied that, at a given age, the RGBB luminosity increases with helium abundance and decreases with metallicity \citep[e.g., see][]{cassisi97}. In the GC MP study, accurate differential photometry can be attainable and the precise estimation of the relative helium abundances between MPs with prior metallicity information can be highly feasible \citep[e.g.,][]{bragaglia10, lee15, lee17, lee24, lagioia18}.

In Figure~\ref{fig:rgbb}, we show the $by5$ CMDs of the RGBB regimes, where the $by5$ index is defined to be $by5 = 5\times(b-y)-2.25$ for the clarity of the figure.\footnote{The reason for using the $by5$ is to expand the $(b-y)$ color by five times so that the separation of individual RGB stars in the CMD can be more clearly seen in the figure. Note that it has nothing to do with the RGBB $V$ magnitude derivation.} We also show the differential luminosity functions (LFs) using a Gaussian kernel density estimation for each population. We calculated the peak values as our RGBB $V$ magnitudes and obtained the RGBB $V$ magnitude for the whole RGB population, \vbump\ = 18.656 $\pm$ 0.030 mag. Our result is in excellent agreement with that of \citet{testa95} who obtained \vbump\ = 18.7 $\pm$ 0.1 mag. Again, this also suggests that our photometric zero-points for NGC\,2257 are reliable.

Unlike Galactic GCs that we previously studied, both the FG and SG of NGC\,2257 show almost identical RGBB $V$ magnitudes; \vbump\ = 18.660 $\pm$ 0.030 mag for the FG and 18.650 $\pm$ 0.035 mag for the SG.  As we will discuss below, the FG and SG have the almost identical metallicity. Therefore, the negligibly small difference in their RGBB $V$ magnitudes in NGC\,2257 suggests that they may have identical helium abundances.
We note that the number of RGB stars near the RGBB in NGC\,2257 is small, which hinders the derivation of accurate RGBB $V$ magnitudes. At the same time, the prominence of the RGBB features depends on metallicity, in the sense that the RGBB feature becomes more indistinguishable with decreasing metallicity \citep[e.g., see][]{fusipecci90, valcarce12}.

We note that \citet{lagioia19a} studied four intermediate age SMC GCs, finding that the degree of the helium enhancement in the SG is relatively small compared to those in the Galactic GCs. Especially, Linsay\,1 does not appear to show the helium enhancement in the SG, while it definitely exhibits CN variations between MPs \citep{niederhofer17b}, similar to NGC\,2257.

We will revisit the identical helium abundance issue with the surface carbon depletion rates, which may provide an alternative way to assess helium abundances.

%%%%%%%%%%%%%%%%%%%%%%%%%%%%%%%%%%%%%%%%%%%%%%%%%%%%%%%%%%%%%%
\begin{figure*}
\epsscale{0.9}
\figurenum{8}
\plotone{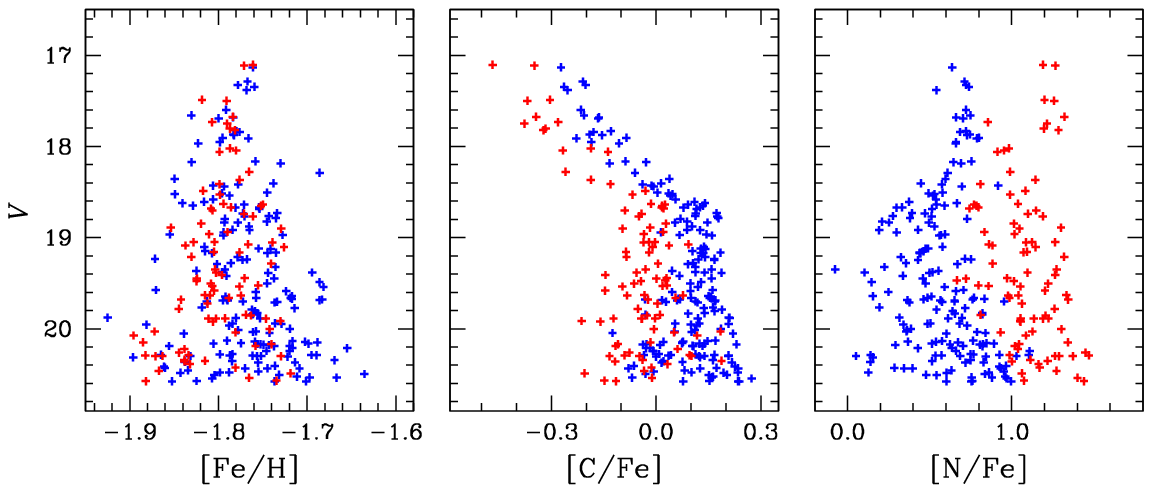}
\caption{The \fehhk, \cfech, and \nfecn\ against $V$ magnitude.
}\label{fig:abund}
\end{figure*}
%%%%%%%%%%%%%%%%%%%%%%%%%%%%%%%%%%%%%%%%%%%%%%%%%%%%%%%%%%%%%%

%%%%%%%%%%%%%%%%%%%%%%%%%%%%%%%%%%%%%%%%%%%%%%%%%%%%%%%%%%%%%%
\begin{figure}
\epsscale{1.2}
\figurenum{9}
\plotone{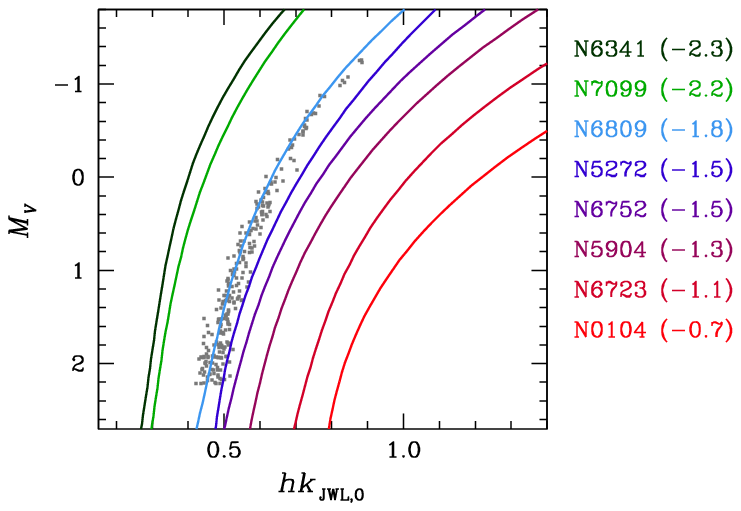}
\caption{The \hkjwl$_{\rm ,0}$ vs.\ $M_V$ CMDs of NGC\,2257 RGB stars, assuming \dmz\ = 18.376 mag and \ebv\ = 0.038 mag for NGC\,2257. We also show fiducial sequences of the eight Galactic GCs with various metallicities from our previous studies.}\label{fig:rgb}
\end{figure}
%%%%%%%%%%%%%%%%%%%%%%%%%%%%%%%%%%%%%%%%%%%%%%%%%%%%%%%%%%%%%%

%===================================
\begin{deluxetable*}{lrrr}
\tablenum{1}
\tablecaption{Elemental Abundances of RGB Stars in NGC\,2257} \label{tab:ab}
\tablewidth{0pc}
\tablehead{\colhead{} & \colhead{\fehhk} & \colhead{\cfech\tablenotemark{a}} & \colhead{\nfecn\tablenotemark{a}}
}
\startdata
All RGB\tablenotemark{b} &  $-$1.777$\pm$0.004($\pm$0.052) & 0.062$\pm$0.007($\pm$0.101) & 0.750$\pm$0.024($\pm$0.345) \\
FG RGB                   &  $-$1.765$\pm$0.004($\pm$0.052) & 0.109$\pm$0.007($\pm$0.076) & 0.566$\pm$0.022($\pm$0.252) \\
 SG RGB\tablenotemark{b} &  $-$1.802$\pm$0.005($\pm$0.044) & $-$0.029$\pm$0.010($\pm$0.079) & 1.115$\pm$0.021($\pm$0.173) \\
\hline
All RGB\tablenotemark{c} &  $-$1.778$\pm$0.004($\pm$0.052) & 0.069$\pm$0.007($\pm$0.095) & 0.742$\pm$0.024($\pm$0.337) \\
 SG RGB\tablenotemark{c} &  $-$1.803$\pm$0.005($\pm$0.043) & $-$0.010$\pm$0.009($\pm$0.078) & 1.091$\pm$0.021($\pm$0.174) \\
\enddata
\tablenotetext{a}{For RGB stars fainter than the RGBB $V$ magnitude.}
\tablenotetext{b}{No helium enhancement in the SG}
\tablenotetext{c}{The helium enhancement of \dy\ = 0.027 in the SG.}
\end{deluxetable*}
%===================================

\subsection{Metallicity}
We derive photometric \feh, \cfe, and \nfe\ of the NGC\,2257 RGB stars using the method that we upgraded in our previous study \citep{lee24}. First, we calculate photometric metallicity, \fehhk, of individual RGB stars. We retrieved the model isochrones for [Fe/H] = $-$2.1, $-$1.9, $-$1.7, and $-$1.5 dex, $Y$ = 0.248, 0.275, and 0.300 with [$\alpha$/Fe] = +0.4 dex, and the age of 11.8 Gyr \citep{milone23} from BaSTI. We note that a change of age by 1 Gyr does not make no larger than 0.03 dex in our current photometric metallicity of the cluster.

We adopted different CNO abundances, [C/Fe] = ($-$0.9, $\Delta$[C/Fe] = 0.3, 0.9), [N/Fe] = ($-$1.5, $\Delta$[N/Fe] = 0.5, 1.5), and [O/Fe] = (0.1, 0.3, 0.5) for each model grid. We constructed 97 model atmospheres and synthetic spectra for each chemical composition from the lower main sequences to the tip of RGB sequences using \atlas\ \citep{kurucz11}, the latest versions of \moogscat\ \citep{sneden74, moogscat}, and \linemake\ \citep{linemake}. As discussed in our previous studies \citep{lee21a, lee23a, lee24}, the RSNH in the metal-poor GC RGB stars is the dominant source of the continuum opacity in the blue and ultra-violet wavelength regions. \moogscat\ takes proper care of RSNH with a non-local thermodynamic equilibrium (non-LTE) treatment of the source function. Following the method described by \citet{giradi02}, we calculated the synthetic color indices for each grid.

To calculate the photometric metallicity, we assumed the following relation,
\begin{eqnarray}
{\rm [Fe/H]}_{hk} &\approx& f_1({{hk}}_{\mathrm{JWL,0}},~Y,~\mathrm{[C,N,O/Fe]},~M_V).\label{eq:FeH}
\end{eqnarray}
We took the same CNO and helium abundances for all RGB stars in the beginning. Using the distance modulus and interstellar reddening value that we derived above, we calculated the $M_V$ and dereddened \hkjwl\ of individual RGB stars. Then we interpolated our synthetic \hkjwl\ index from fine model grids to match with the observed $hk_{\rm JWL,0}$ index.
In a recursive manner, the input carbon and nitrogen abundances are updated using the \cfech, the carbon abundance from our \chjwl\ index, and \nfecn, the nitrogen abundance from our \cnjwl\ index, by using the relations given below. Since we do not have oxygen abundances of individual stars, we assigned [O/Fe] = 0.5 dex for the FG and 0.1 dex for the SG. We note that the presumed oxygen abundance does not significantly affect our results. We iterated these processes until the derived metallicity converges within $\leq$ 0.01 dex.

We obtained the mean \fehhk\ = $-$1.777$\pm$0.004 dex ($\sigma$ = 0.052 dex) for NGC\,2257, without the helium enhancement in the SG. We show our results in Figure~\ref{fig:abund} and Table~\ref{tab:ab}. For individual populations, we obtained \fehhk\ =  $-$1.765$\pm$0.004 dex ($\sigma$ = 0.052 dex) for the FG and $-$1.802$\pm$0.005 dex ($\sigma$ = 0.044 dex) for the SG without the helium enhancement in the SG (see below). We note that the scatter in our photometric metallicity of NGC\,2257 is comparable to those of monometallic GCs, M30 \citep{lee24} and M5 \citep{lee21b}, or those of individual groups in the metal-complex GCs, 47~Tuc \citep{lee22}, M3 \citep{lee21a}, M22 \citep{lee23a} and M92 \citep{lee23b, lee24}, suggesting that NGC\,2257 is a genuine monometallic GC.

As a sanity check, we show dereddened CMDs of the NGC\,2257 RGB stars in Figure~\ref{fig:rgb} along with the RGB fiducial sequences of Galactic GCs that we study previously \citep{lee15, lee17, lee18, lee19b, lee23a, lee24, lee21a}. The NGC\,2257 RGB sequence is in excellent agreement with that of M55 (\feh\ $\sim$ $-$1.8 dex, \citealt{lee16, gontcharov23}). We note that our \hkjwl\ index is insensitive to the interstellar reddening, $E$(\hkjwl) = $-$0.067 \ebv\ \citep[e.g., see][]{lee23b}, and, as a consequence, our photometric metallicity is not vulnerable to the inaccurate interstellar reddening or the differential reddening effect across our science field.

As we already discussed for the same RGBB $V$ magnitudes, NGC\,2257 does not appear to show any helium abundance difference between the FG and SG. Nonetheless, we calculated the metallicity for the SG population applying the helium enhancement of \dy\ = 0.027 for the SG, which is about the mean helium enhancement of the SG with respect to the FG in Galactic normal GCs in our previous studies, M5 and M30 \citep{lee21b, lee24}, in order to demonstrate how the helium abundance influences our metallicity derivations.  We obtained \fehhk\ = $-$1.803$\pm$0.005 dex ($\sigma$ = 0.043 dex) for the SG with an enhanced helium abundance and it is almost identical to that without the helium enhancement. It strongly suggests that our adopted helium abundance of the SG does not significantly affect our \fehhk\ measurements.

It is worrisome that our mean \fehhk\ value is about 0.15 -- 0.20 dex more metal rich than that reported by \citet{mucciarelli10}, \feh\ = $-$1.95$\pm$0.02 dex ($\sigma$ = 0.04 dex). They obtained the rather low signal-to-noise ratio high-resolution ($\lambda/\Delta\lambda \sim$ 45,000) and intermediate-resolution ($\lambda/\Delta\lambda \sim$ 24,000) spectra of six RGB stars in NGC\,2257.\footnote{Without the coordinates of the target RGB stars studied by \citet{mucciarelli10}, we were not able to compare our \feh\ with their measurements.} \citet{lee24} discussed that the determination of the effective temperature is not a trivial matter in high-resolution spectroscopy and it can significantly mislead the results. We believe that the discrepancy in the metallicity of NGC\,2257 is mainly due to incorrect temperature assignments by \citet{mucciarelli10}.

\begin{figure}
\epsscale{1.1}
\figurenum{10}
\plotone{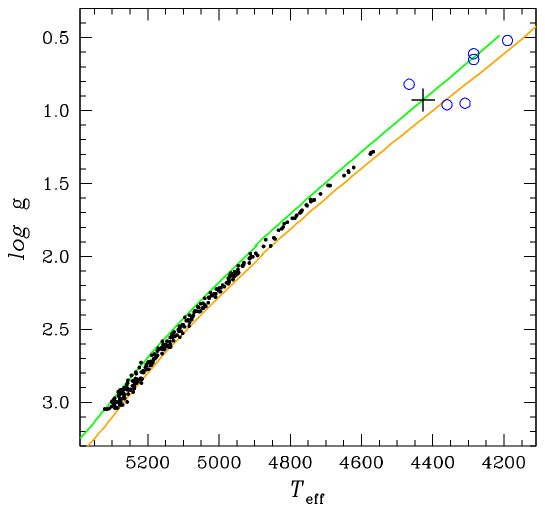}
\caption{A plot of \teff\ vs.\ $\log g$ for NGC\,2257 RGB stars from our evolutionary populational model (black dots). The blue open circles denote the stellar parameters of RGB stars derived by \citet{mucciarelli10}. The thick green and orange lines are for the BaSTI model RGB sequence for [Fe/H] = $-$1.90 and $-$1.70 dex \citep{basti21}. The black plus sign denotes the location of the observed RGB tip of NGC\,2257 ($V$ $\sim$ 16.5 mag) for [Fe/H] = $-$1.90 dex.
}\label{fig:pam}
\end{figure}

In Figure~\ref{fig:pam}, we show a plot of \teff\ versus $\log g$ of the NGC\,2257 RGB stars used in our study, along with those of the model isochrones for [Fe/H] = $-$1.90 and $-$1.70 \citep{basti21}. Note that the brightest RGB stars in NGC\,2257 has $V$ $\sim$ 16.5 mag (i.e., the \emph{observed} RGB-tip $V$ magnitude) as shown in Figure~\ref{fig:cmd} (see also \citealt{walker89} and \citealt{testa95}).
The brightest RGB star studied by \citet{mucciarelli10} is NGC\,2257-993, whose dereddened $K$ magnitude is $K_0$ = 13.49 mag. Using the distance modulus of \dmz\ = 18.37 mag \citep[][and our current work]{milone23}, the absolute $K$ magnitude becomes $M_K$ = $-$4.88 mag. Then the effective temperature from a BaSTI model for [Fe/H] = $-$1.90 dex and 11.8 Gyr is about 4410 K,\footnote{The small difference in the adopted age does not affect our results presented here. For the age of 12.5 Gyr, the effective temperature for $M_K$ = $-$4.88 mag is about 4400 K.}  in sharp contrast to the photometric effective temperature, $T_{\rm eff}^{\rm phot}$ = 4190 K, adopted by \citet{mucciarelli10}. In fact, their effective temperature is very close to the isochrone's effective temperature at the RGB-tip (not the \emph{observed} RGB-tip of NGC\,2257), $\sim$ 4194 K. We suspect that \citet{mucciarelli10} mistakenly put the star NGC\,2257-993 at the RGB-tip of their adopted stellar models. Same is true for remaining five stars: The effective temperature adopted by \citet{mucciarelli10} is about 200 K cooler than our evolutionary effective temperature, which makes the mean \feh\ of \citet{mucciarelli10} is about 0.15 -- 0.2 dex more metal-poor than our measurements \citep[e.g., see][$\delta$\feh/$\delta$\teff($-$100 K) $\sim$ $-$0.1 dex at \feh\ $\sim$ $-$1.8 dex]{lee02}.

%%%%%%%%%%%%%%%%%%%%%%%%%%%%%%%%%%%%%%%%%%%%%%%%%%%%%%%%%%%%%%
%%%%%%%%%%%%%%%%%%%%%%%%%%%%%%%%%%%%%%%%%%%%%%%%%%%%%%%%%%%%%%

%%%%%%%%%%%%%%%%%%%%%%%%%%%%%%%%%%%%%%%%%%%%%%%%%%%%%%%%%%%%%%
\begin{figure}
\epsscale{1.2}
\figurenum{11}
\plotone{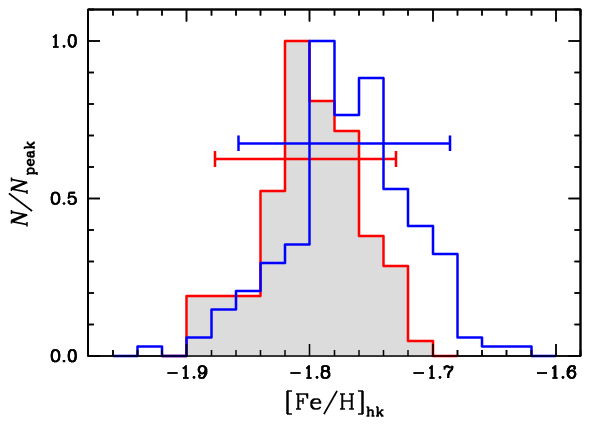}
\caption{The \fehhk\ histograms for the FG (the blue solid line) and SG (the red solid line shaded with gray) in NGC\,2257. The small vertical lines denote the locations of the 5th and 95th percentiles of \fehhk\ distributions for each population.}\label{fig:feh}
\end{figure}
%%%%%%%%%%%%%%%%%%%%%%%%%%%%%%%%%%%%%%%%%%%%%%%%%%%%%%%%%%%%%%

%%%%%%%%%%%%%%%%%%%%%%%%%%%%%%%%%%%%%%%%%%%%%%%%%%%%%%%%%%%%%%
\begin{figure}
\epsscale{1.2}
\figurenum{12}
\plotone{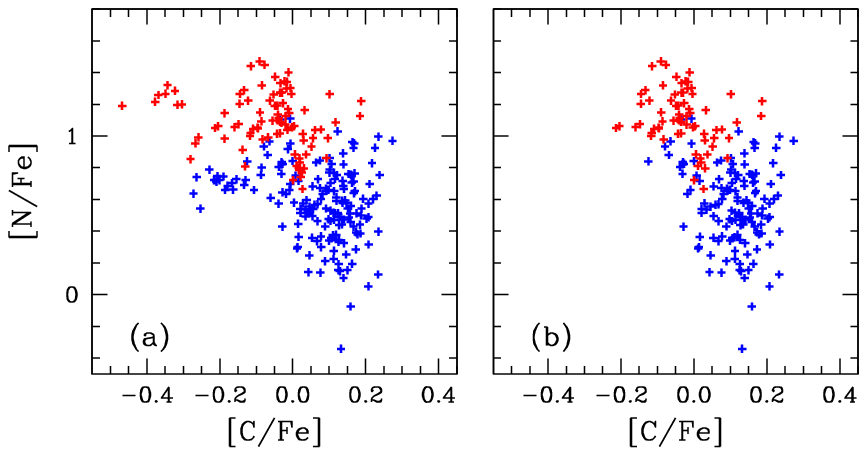}
\caption{(a) A plot of the \cfe\ vs. \nfe\ for all RGB stars in NGC\,2257.
(b)  A plot of the \cfe\ vs. \nfe\ for all RGB stars fainter than RGBBs.
}\label{fig:cfenfe}
\end{figure}
%%%%%%%%%%%%%%%%%%%%%%%%%%%%%%%%%%%%%%%%%%%%%%%%%%%%%%%%%%%%%%

\subsection{Metallicity Variations in the FG and SG}
By employing a high precision differential line-by-line analysis of high-quality spectra of the NGC\,6752 RGB stars, \citet{yong13} clearly showed that the \feh\ variation of the FG is much larger than that of the SG. In recent study, \citet{legnardi22} argued that the metallicity variations of the FG in the 55 Galactic GCs range from \dfehfg\ of less than 0.05 dex to 0.30 dex, while the SG stars have more homogeneous metallicity distributions than the FG. They also suggested that the metallicity variations are mildly correlated with cluster mass.

As we already showed in Table~\ref{tab:ab}, the \fehhk\ dispersion of the FG, $\sigma$ = 0.052, is larger than that of the SG, $\sigma$ = 0.044. In Figure~\ref{fig:feh}, we show the \fehhk\ histograms for the FG and SG in NGC\,2257. We also show the locations of the 5th and 95th percentiles, and the widths of the \fehhk\ distributions, $\delta$\feh, which are defined to be \fehhk(95th) minus \fehhk(5th), for each population.
We obtained \dfehfg\ = 0.172 dex and \dfehsg\ = 0.147 dex. The metallicity distribution of the NGC\,2257 SG stars is narrower than the FG, similar to findings for NGC\,6362 and NGC\,6838 by \citet{legnardi22}. Our result represents the first detection of such phenomenon in an old LMC GC with MPs and, furthermore, the larger metallicity variations in the FG could be a general characteristic of GCs with MPs, not restricted to the Galactic GCs.

Finally, we note that the widths are the same as 3.310$\times\sigma$, meaning that the \fehhk\ distributions of each population in NGC\,2257 follow the Gaussian distribution. On the other hand, the pseudo-color and metallicity distributions of the FG RGB stars in NGC\,6362 and NGC\,6838 are non-Gaussian, which may have rooted in the different mechanism than in NGC\,2257.

\subsection{Carbon and Nitrogen Abundances}
The photometric carbon and nitrogen abundances are estimated using the following relations,
\begin{eqnarray}
{\rm [C/Fe]}_{ch} &\approx& f_2({{ch}}_{\mathrm{JWL,0}},{\mathrm{[Fe/H]}_{hk}},{\rm [N/Fe]}_{cn},Y,\mathrm{[O/Fe]},M_V), \\
{\rm [N/Fe]}_{cn} &\approx& f_3({{cn}}_{\mathrm{JWL,0}},{\mathrm{[Fe/H]}_{hk}},{\rm [C/Fe]}_{ch},Y,\mathrm{[O/Fe]},M_V).
\end{eqnarray}
As mentioned above, we were not able to obtain the JWL34 images due to the UV insensitive property of the Gemini-South 8.0 m telescope and the GMOS combination. Instead, we derive the nitrogen abundance from our \cnjwl\ measurements \citep[see also][]{lee21a}. Similar to our photometric metallicity, our photometric carbon and nitrogen abundances are not vulnerable to our adopted interstellar reddening value or the differential reddening effect: $E$(\chjwl) = $-$0.043 \ebv\ and $E$(\cnjwl) = 0.053 \ebv\ \citep{lee23b}.

In Figure~\ref{fig:abund}, we show \cfech\ and \nfecn\ against the $V$ magnitude. The figure clearly exhibits the hallmark of internal deep mixing episodes accompanied by the CN cycle in bright RGB stars, which induce the surface carbon depletion and the nitrogen enhancement with the RGB $V$ magnitude \citep[see][and references therein]{lee23b}. In Table~\ref{tab:ab}, we show the mean \cfech\ and \nfecn\ values for the RGB stars fainter than their RGBB $V$ magnitudes. We show a plot of \cfech\ versus \nfecn\ in Figure~\ref{fig:cfenfe}, showing that NGC\,2257 exhibits a C--N anticorrelation that can be seen in the Galactic normal GCs.

%%%%%%%%%%%%%%%%%%%%%%%%%%%%%%%%%%%%%%%%%%%%%%%%%%%%%%%%%%%%%%
\begin{figure}
\epsscale{1.15}
\figurenum{13}
\plotone{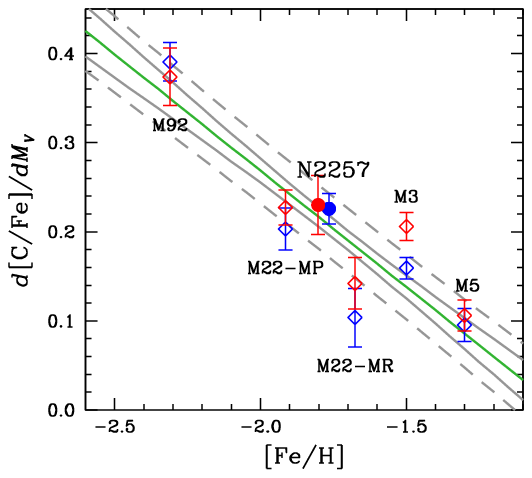}
\caption{A plot of the \cfemv\ vs.\ \feh\ of NGC\,2257 and four Galactic GCs from our previous study \citep{lee23b}. The blue and red colors denote the FG and SG, respectively. The green line shows a linear regression, while the grey solid and dashed lines denote 95\% confidence and predicted intervals.}\label{fig:grad}
\end{figure}
%%%%%%%%%%%%%%%%%%%%%%%%%%%%%%%%%%%%%%%%%%%%%%%%%%%%%%%%%%%%%%

\subsection{Surface Carbon Depletion Rates as a Robust Probe of Relative He Abundances between MPs}
As we discussed already, due to internal mixing episodes accompanied by the CN cycle in the hydrogen shell burning region, the surface carbon and nitrogen abundances of the RGB stars brighter than their RGBB are significantly changed from their initial abundances \citep[e.g.,][]{sneden86, charbonnel07}. Furthermore, the mixing efficiency sensitively depends on metallicity and helium abundance since they effectively alter the internal temperature structures of the low-mass RGB stars \citep[e.g., see][]{church14}.

In our previous study \citep{lee23b}, we demonstrated that the surface carbon depletion rates (CDRs) against the absolute magnitude of four Galactic GCs are nicely correlated with their metallicities,
\begin{equation}
 \frac{d\mathrm{[C/Fe]}}{dM_V} = -0.261(\pm0.043)\mathrm{[Fe/H]} - 0.253(\pm0.077).\end{equation}
In addition, \citet{lee23b} argued that the SG of Galactic GCs have slightly steeper CDRs than the FG does, due to enhanced helium abundance which induce higher central temperature. As a consequence, the SG stars brighter than the RGBB with enhanced helium abundance may experience faster carbon destruction due to temperature dependency in the CN cycle than the FG.

Following the same method that we employed in \citet{lee23b}, we derived the CDRs for the FG and SG of NGC\,2257 RGB stars brighter than their RGBBs. We obtained the identical CDRs for both populations assuming no helium enhancement in the \cfech\ derivations for the SG, \cfemv\ = 0.226($\pm$0.017) dex mag$^{-1}$ for the FG and  0.230($\pm$0.033) dex mag$^{-1}$ for the SG. We also obtained 0.227($\pm$0.032) dex mag$^{-1}$ assuming a potential helium enhancement of \dy\ = 0.027 in the derivations of \cfech\ for the SG, which is nearly identical to that without the helium enhancement. We show our results in Figure~\ref{fig:grad}. As shown in the figure, the CDRs of NGC\,2257 are in excellent agreement with our previous relation from Galactic GCs. It is not a surprise, since the CDR is mainly governed by the stellar internal structures, not by the physical environments: The internal structure of the low-mass stars in NGC\,2257 are similar to those in the Galactic GCs with similar  metallicities. We also note that the identical CDRs between the two populations in NGC\,2257 may support the idea that there is no significant helium variation among different populations.

As we discussed above, due to the small number of RGB stars in NGC\,2257, the derivation of the relative helium abundances between the two populations by comparing their RGBB $V$ magnitudes has some limitation and may not be precise enough to investigate the helium enhancement in the SG. Alternatively, the degree of the CDR can provide a robust independent test to examine the difference in the internal stellar structures governed mostly by helium abundance of individual MPs with the same metallicity.

We conclude that there is no substantial observational line of evidence for the helium enhancement in the SG with respect to the FG in NGC\,2257.

%%%%%%%%%%%%%%%%%%%%%%%%%%%%%%%%%%%%%%%%%%%%%%%%%%%%%%%%%%%%%%
\begin{figure}
\epsscale{1.1}
\figurenum{14}
\plotone{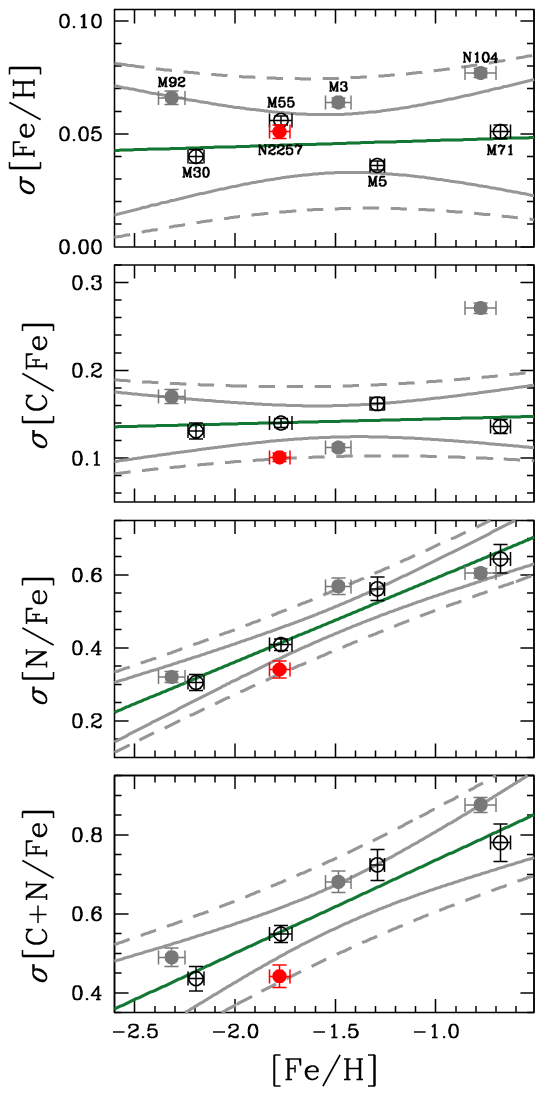}
\caption{Plots of the \feh\ vs. \sfeh, \scfe, \snfe, and \scpnfe\ for NGC\,2257 (a filled red circle), four Galactic normal GCs (open circles), and three metal-complex GCs (grey filled circles). The thick green lines denotes the linear regressions of the four Galactic normal GCs, while the thick grey and dashed grey lines denote the 95\% confidence and predicted intervals.
}\label{fig:dispersion}
\end{figure}
%%%%%%%%%%%%%%%%%%%%%%%%%%%%%%%%%%%%%%%%%%%%%%%%%%%%%%%%%%%%%%
%%%%%%%%%%%%%%%%%%%%%%%%%%%%%%%%%%%%%%%%%%%%%%%%%%%%%%%%%%%%%%
\begin{figure}
\epsscale{1.1}
\figurenum{15}
\plotone{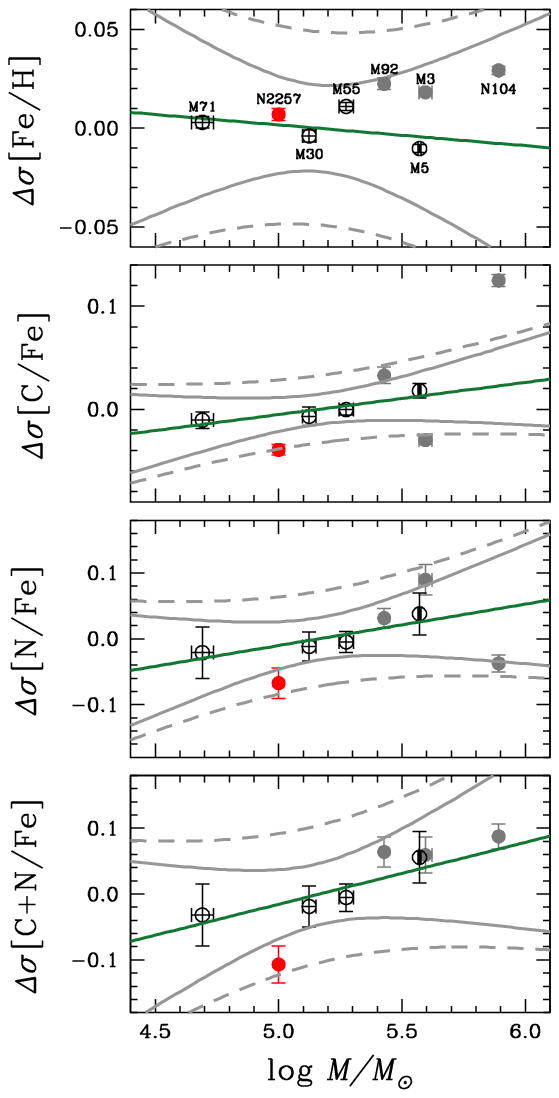}
\caption{Same as Figure~\ref{fig:dispersion} but for the $\log M/M_{\odot}$ vs. $\Delta$\sfeh, $\Delta$\scfe, $\Delta$\snfe, and $\Delta$\scpnfe.
}\label{fig:mass}
\end{figure}
%%%%%%%%%%%%%%%%%%%%%%%%%%%%%%%%%%%%%%%%%%%%%%%%%%%%%%%%%%%%%%

\subsection{$\sigma${\rm[C/Fe]} and $\sigma${\rm[N/Fe]}}
In sharp contrast to broad-band pseudo color indices widely being used, which involve combined effects from different diatomic molecules, such as OH, NH, CN, and CH, with different degrees of luminosity effects \citep[e.g., see][]{lee19a, lee23a, lee24}, the strong point of our photometric system is that we are able to directly measure carbon and nitrogen abundances of individual RGB stars in GCs. For carbon and nitrogen, it is important to obtain their abundances from stars fainter than the RGBB, where primordial carbon and nitrogen abundances are attainable. Therefore, we can clearly delineate the difference and evolution of the carbon and nitrogen abundances among GCs. Here, we look more into the carbon and nitrogen abundance dispersions.

In Figure~\ref{fig:dispersion}, we show plots of the \feh\ versus \sfeh, \scfe, \snfe, and \scpnfe\ of RGB stars fainter than their RGBBs in NGC\,2257 and some Galactic GCs including those from our previous studies \citep{lee15, lee17, lee18, lee19b, lee23a, lee24, lee21a}.\footnote{Note that [C,N/Fe] of RGB stars in NGC\,104, M55, and M71 are not published previously. We made observations for these clusters using the WIYN 0.9 m and the SWOPE 1 m telescopes and we will present detailed discussions for these clusters in our forthcoming papers.} In the figure, the metal-complex GCs (NGC\,104, M3, and M92 in the order of decreasing \feh) show different relations from those from the Galactic normal GCs (M71, M5, M55, and M30). We point out a strong  \snfe--\feh\ correlation among Galactic normal GCs, which is consistent with what \citet{milone17} found. They showed that the intrinsic RGB color width are nicely correlated with [Fe/H] in Galactic GCs. In particular, the increase in the nitrogen abundance variation affects the formation of diatomic molecules, such as NH and CN, with increasing metallicity, which naturally leads to broader pseudo-color indices that \citet{milone17} adopted.
In our following analyses, we do not rely on these metal-complex GCs to derive any relations.

\subsection{$\Delta\sigma${\rm[C/Fe]} and $\Delta\sigma${\rm[N/Fe]} versus $N_{\rm FG}/N_{\rm tot}$}
Figure~\ref{fig:dispersion} clearly shows that there exist metallicity effects on elemental abundance dispersions, in particular for \snfe\ and \scpnfe. To remove these metallicity effects in the elemental abundance dispersions of the Galactic GCs, we follow the similar procedure that \citet{milone17} did. We find the best fitted lines in the \sfecnfe\ versus \feh\ relations, which are shown with thick green solid lines in the figure. Then we derive the metallicity dependency removed abundance variations, \dsfecnfe. We show plots of the \dsfecnfe\ versus present-day GC masses in Figure~\ref{fig:mass}. The slope of the  \dsfeh\ versus \logmass\ relation is almost nil for the Galactic normal GCs, since our measurement of the metallicity dispersions, \sfeh, in the Galactic normal GCs are almost the same regardless of metallicity, \sfeh\ $\sim$ 0.04--0.05 dex, as already shown in Figure~\ref{fig:dispersion}.

The plots of the \dscnfe\ and \dscpnfe\ versus \logmass\ of the Galactic normal GCs show tight positive correlations, meaning that both the carbon and nitrogen abundance variations increase with the present-day GC masses. Our results suggest that the more massive Galactic GCs might experience more active chemical evolution history at a given metallicity. On the other hand,  the metal-complex GCs are distributed rather far from the fitted lines, suggesting that they may have different formation or evolution histories from the normal GCs. We note that our results are consistent with those of others \citep{milone17, lagioia19b, dondoglio2021}, who argued that the metallicity removed intrinsic width of the RGB pseudo color indices is positively correlated with the GC mass. Although the number of our samples is very limited, the LMC GC NGC\,2257 exhibits slightly weaker \dscnfe\ than the Galactic normal GCs with the similar mass. Assuming the \dscnfe\ are pseudo-invariants, our results may suggest that NGC\,2257 and Galactic normal GCs experience different chemical evolutions, or, more likely, the GCs in our Galaxy experience more active mass-loss over the Hubble time as \citet{lagioia19b} suggested.

%%%%%%%%%%%%%%%%%%%%%%%%%%%%%%%%%%%%%%%%%%%%%%%%%%%%%%%%%%%%%%
\begin{figure*}
\epsscale{1.}
\figurenum{16}
\plotone{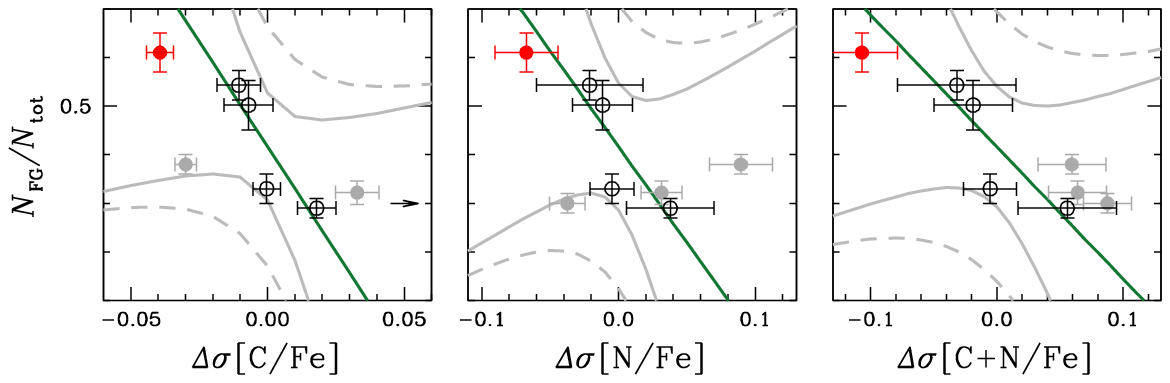}
\caption{Plots of the \dscnfe\ vs. \fgfrac\ for four Galactic normal GCs. The green lines show linear regressions, while the grey solid and dashed lines denote 95\% confidence and predicted intervals. The red filled circles denote NGC\,2257, which is in good agreement with Galactic normal GCs.
}\label{fig:ci}
\end{figure*}
%%%%%%%%%%%%%%%%%%%%%%%%%%%%%%%%%%%%%%%%%%%%%%%%%%%%%%%%%%%%%%

%%%%%%%%%%%%%%%%%%%%%%%%%%%%%%%%%%%%%%%%%%%%%%%%%%%%%%%%%%%%%%
\begin{figure*}
\epsscale{0.9}
\figurenum{17}
\plotone{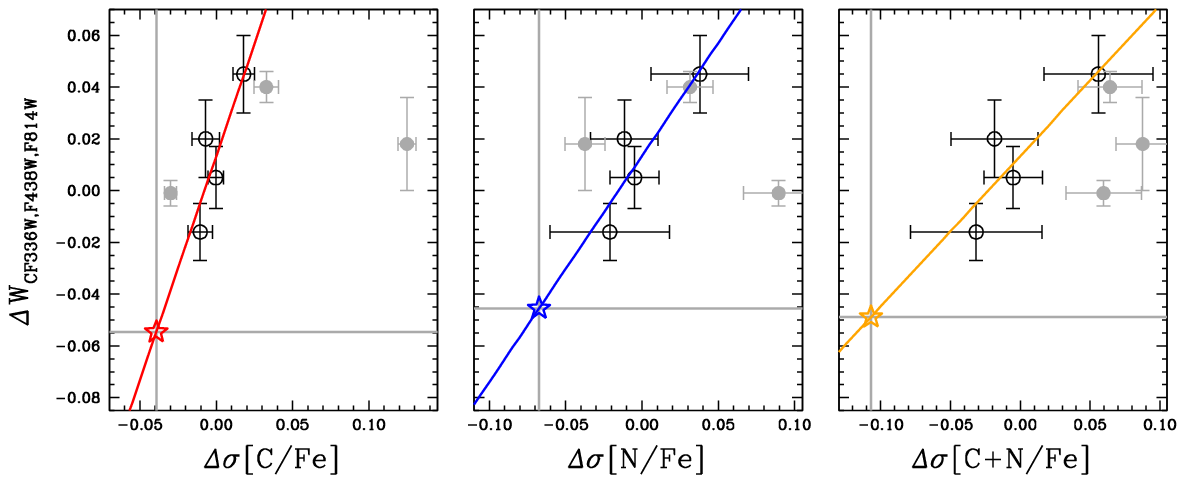}
\caption{Plots of \dscnfe\ vs. \rgbw\ of four Galactic normal GCs from Figure~\ref{fig:ci}. The red, blue, and orange open stars denote the predicted \rgbw\ of NGC\,2257.
}\label{fig:cvt}
\end{figure*}
%%%%%%%%%%%%%%%%%%%%%%%%%%%%%%%%%%%%%%%%%%%%%%%%%%%%%%%%%%%%%%

We believe that the \dscnfe\ of the RGB stars fainter than the RGBB can be a pseudo--invariant parameter for normal GC MPs, since the primordial carbon and nitrogen abundance dispersions set at the formation epochs of individual populations. Furthermore, the stellar atmospheres of the RGB stars fainter than the RGBB retain their pristine chemical compositions.\footnote{We neglect the influence of binaries, which can significantly alter the surface carbon abundances \citep[e.g.,][]{starkenburg14}.} On the other hand, the present-day GC mass, for example, could not be an invariant parameter, although their present-day masses may reflect their initial masses in some part. In the context of the GC MP evolution, a significant fraction of FG stars (and the SG stars with a less significant degree, as well) must have been lost to be assembled to the host galaxy in the early evolution of GCs \citep{dercole08, ventura14, vesperini21}. Also numerous stellar streams associated with Galactic GCs vividly demonstrate the secular mass loss due to tidal disruptions of GCs \citep[e.g.,][]{helmi20}.

Instead of relying on the present-day GC masses, we investigate the \dscnfe, \dscpnfe\ versus \fgfrac, as shown in Figure~\ref{fig:ci}. In the figure, we also show the linear regressions with 95\% confidence and predicted intervals from the four Galactic normal GCs. Although a small number of our sample size, the metallicity dependency removed carbon and nitrogen abundance dispersions, \dscnfe, appear to be tightly correlated to the populational number ratios, \fgfrac\ : The fraction of the FG decreases (i.e., the fraction of the SG increases) with  increasing \dscnfe. Furthermore, NGC\,2257 appears to follow these trends.

This behavior could be understood in this way. Numerical simulations done by \citet{vesperini21}, for example, suggested that the MP populational ratio does not change significantly after the first few Gyrs (when GCs enter a dynamical phase driven by the effects of the two-body relaxation) of the GC evolution. Since GCs with large \dscnfe\ values are expected to be formed in a large initial GC mass, our \dscnfe\ can be an alternative measure of the initial GC mass. Therefore, it is believed that \dscnfe, \dscpnfe\ versus \fgfrac\ relations could be alternative expressions for the initial GC mass versus \fgfrac\ relation.

We expand our idea using the HST observations by \citet{milone17,milone20, lagioia19b, vanaraj21}.
Since there is no previous HST F438W observation for NGC\,2257, we attempt to derive \rgbw\ of NGC\,2257 from our \dscnfe\ measurements, assuming that NGC\,2257 is a normal GC. As we mentioned above, it is not an absurd idea since our \dscfe\ and \dsnfe\ should control the intrinsic RGB widths of the pseudo color indices, such as \rgbwo.

In Figure~\ref{fig:cvt}, we show plots of \dscnfe, \dscpnfe\ versus \rgbw\ for Galactic normal GCs, exhibiting tight correlations. From these relations, we obtained the inferred \rgbw\ of 0.050 $\pm$ 0.046 for NGC\,2257 from our \dscnfe\ measurements for NGC\,2257.

\begin{figure*}
\epsscale{1.0}
\figurenum{18}
\plotone{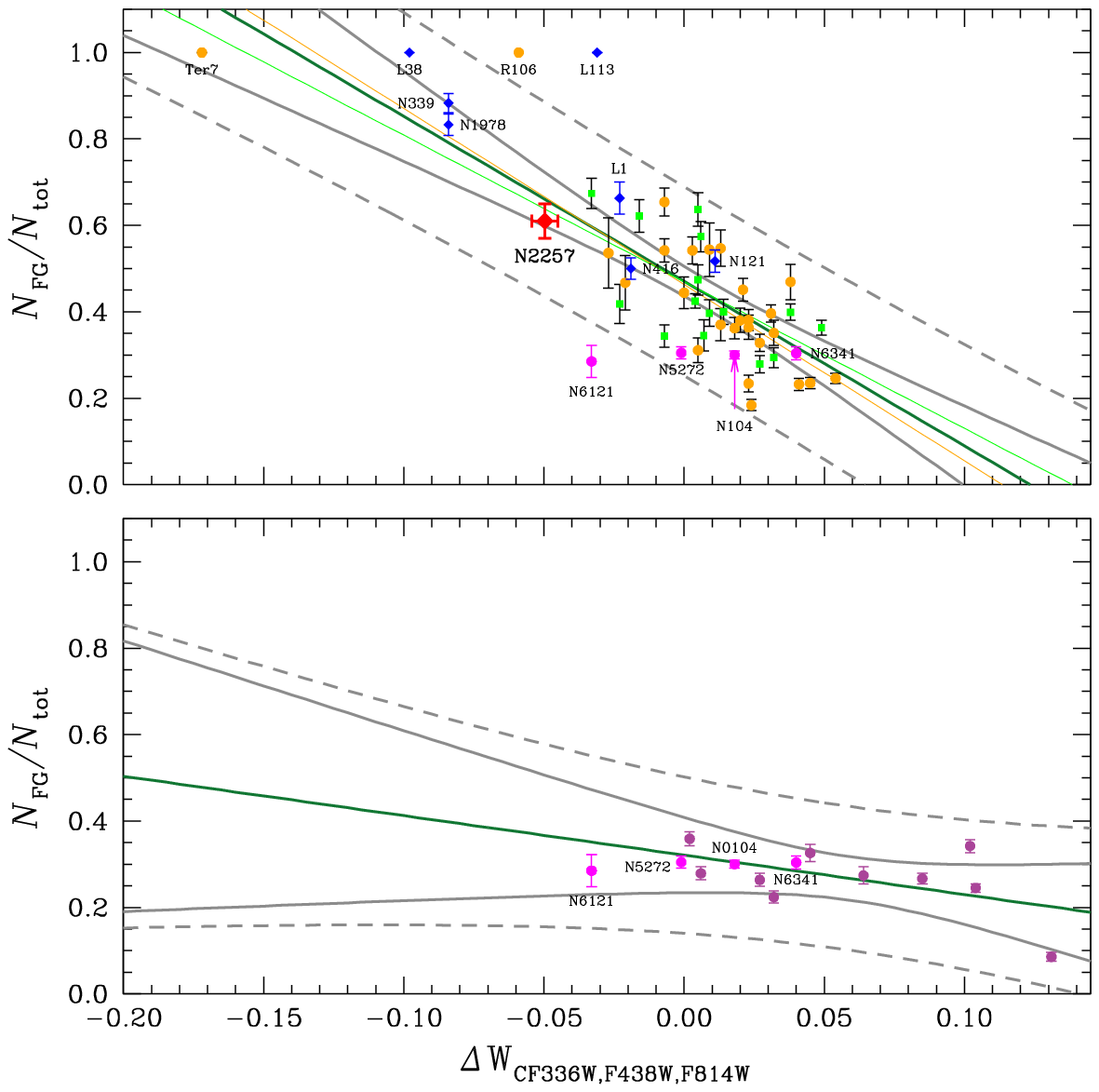}
\caption{(Top) A plot of \rgbw\ vs. \fgfrac. The green filled boxes denote the in situ Galactic GCs, the orange filled circles the ex situ Galactic GCs, the blue filled diamonds the young MC GCs. NGC\,2257 is shown with a big red diamond. The thick dark green line denotes the linear regression from the normal Galactic GCs (both in situ and ex situ), while the solid and dashed grey lines are for 95\% confidence and predicted intervals. The thin green and orange solid lines show a linear regression for in situ and ex situ Galactic GCs, both of which are similar.
(Bottom) A plot of \rgbw\ vs. \fgfrac\ for metal-complex Galactic GCs (red filled circles).  The green line shows a linear regression of the metal-complex GCs, while the grey solid and dashed lines denote 95\% confidence and predicted intervals.
}\label{fig:hst}
\end{figure*}

In Figure~\ref{fig:hst}, we show a plot of the \rgbw\ versus \fgfrac\ of the Galactic and MC GCs, including NGC\,2257. We also show a linear regression from the 41 Galactic normal GCs. Note that we do not include the metal-complex GCs and M4 (see below). The figure suggests that there is a strong anticorrelation with a correlation coefficient of $\rho = -0.750$  between the \rgbw\ and \fgfrac\ among Galactic normal GCs (both the in situ and ex situ) shown with the thick green lines. As we already mentioned, the \rgbw\ could be a good proxy of our \dscnfe\ and \dscpnfe.

We would like to point out seven important aspects on this plot:
\begin{itemize}
\item[(1)] NGC\,2257 is in excellent agreement with the relation from the Galactic normal GCs.
\item[(2)] The \fgfrac\ of NGC\,104 should be 0.30$\pm$0.01 \citep[shown with an arrow in the figure;][]{lee22}. NGC\,104 shows a very strong radial gradient in the populational number ratio, in the sense that the SG is more centrally concentrated. Due to a very small field of view of the HST observation, \citet{milone17} obtained a small value of \fgfrac\ (= 0.175$\pm$0.009) for the central part of the cluster.
\item[(3)] M4 (NGC\,6121, see below) and Ruprecht\,106 deviate significantly from the fitted line.
\item[(4)] Both the in situ and ex situ Galactic GCs appear to have the same trend. It is natural to expect that the degree of tidal disruption of the GC MPs, which will reflect the current populational number ratios and total masses, could be different between the in situ and ex situ GCs. The same trend between the in situ and ex situ Galactic GCs is most likely due to the fact that the metallicity removed abundance dispersions or the RGB width may well reflect the initial GC mass as we argued above.
\item[(5)] The metal-complex GCs classified by \citet{milone17} (the filled purple circles) and by us (the filled magenta circles; NGC\,104, M3, and M92) in the bottom panel of Figure~\ref{fig:hst} appear to have a distinct \rgbw\ versus \fgfrac\ relation from the Galactic normal GCs. We suspect that M4 could be one of such GCs. These GCs mush have very different MP formation histories than the Galactic normal GCs.
\item[(6)] The most of the MC GCs follow the anticorrelation set by the Galactic normal GCs. One outlier, Lindsay\,113, is a very young GC ($\sim$ 3.6 Gyr, \citealt{milone23}), and it may not be fair to be compared with rest of old GCs.
\item[(7)] Terzan\,7 is reported to have a simple stellar population \citep{lagioia24} and it appears to follow the same trend as the Galactic normal GCs.
\end{itemize}

\begin{figure*}
\epsscale{1.0}
\figurenum{19}
\plotone{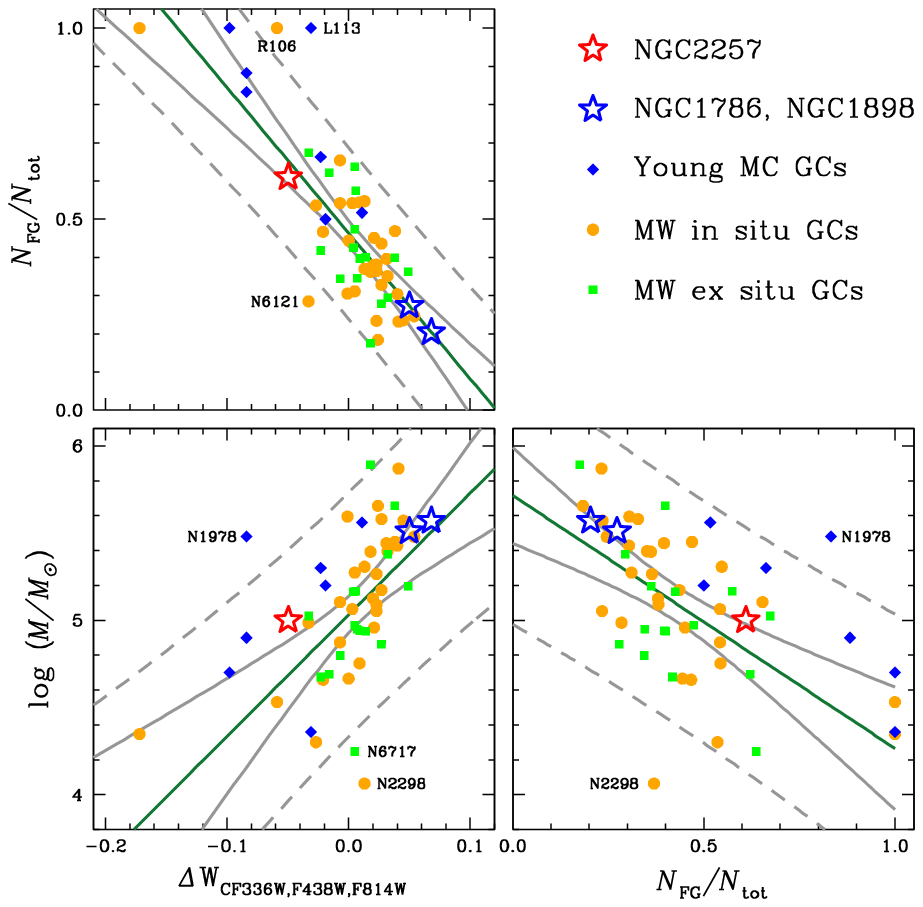}
\caption{Plots of \rgbw\ vs. \logmass\ vs. \fgfrac.
}\label{fig:cpm}
\end{figure*}

Finally, we show plots of \rgbw\ vs. \logmass\ vs. \fgfrac\ of Galactic and MC GCs in Figure~\ref{fig:cpm}. In the figure, we include two old LMC GCs, NGC\, 1786 and NGC\,1898 \citep{vanaraj21}. The \fgfrac\ values of these two GCs are not known. We simply applied the \fgfrac\ versus \rgbw\ relation in Figure~\ref{fig:hst}, assuming these two GCs are normal GCs, and we obtained \fgfrac\ = 0.204 $\pm$ 0.073 and 0.273 $\pm$ 0.053 for NGC\,1786 and NGC\,1898, respectively. Similar what \citet{vanaraj21} argued, three old LMC GCs are in good agreement with Galactic normal GCs in these domains. We conclude that there is no significant difference between the Galactic (both the in situ and ex situ) and old LMC GC MPs, which may indicate that the galaxy environment did not play a major role in the formation and evolution of GC MPs.

%%%%%%%%%%%%%%%%%%%%%%%%%%%%%%%%%%%%%%%%%%%%%%%%%%%%%%%%%%%%%%
%%%%%%%%%%%%%%%%%%%%%%%%%%%%%%%%%%%%%%%%%%%%%%%%%%%%%%%%%%%%%%

%%%%%%%%%%%%%%%%%%%%%%%%%%%%%%%%%%%%%%%%%%%%%%%%%%%%%%%%%%%%%%
%%%%%%%%%%%%%%%%%%%%%%%%%%%%%%%%%%%%%%%%%%%%%%%%%%%%%%%%%%%%%%

\section{SUMMARY}
As part of our ongoing effort to investigate the GC MPs using our own photometric system, we expanded our study to the old metal-poor LMC GC, NGC\,2257. Our distance modulus and interstellar reddening value are in excellent agreement with previous measurements by others.

The strong point of our photometric system is that we can simultaneously measure photometric metallicity, carbon and nitrogen abundances of individual stars. We obtained the mean photometric metallicity of \fehhk\ = $-$1.777$\pm$0.004 dex ($\sigma$ = 0.052). A small metallicity dispersion of NGC\,2257 suggested that it is a genuine mono-metallic GC. Our metallicity for the cluster is in excellent agreement with previous measurement by others. However, our \fehhk\ is about 0.2 dex more metal-rich than that claimed by \citet{mucciarelli10}, $-$1.95$\pm$0.02 dex ($\sigma$ = 0.04). We showed that this large difference is due to significantly low temperature scales by $\Delta$\teff\ $\sim -$200 K adopted by \citet{mucciarelli10}.

We showed that the degree of the metallicity variation of the FG is somewhat larger than that of the SG in NGC\,2257, similar to Galactic GCs with MPs.
Our result represents the first detection of such phenomenon in an old LMC GC with MPs, strongly suggesting that it is a general characteristic of GCs with MPs, not restricted within the Galactic GCs. However, the FG metallicity distribution in NGC\,2257 appears to follow the Gaussian distribution, while those of Galactic GCs by \citet{legnardi22} are clearly non-Gaussian, may suggesting different inducing mechanisms between the two.

We derived the photometric carbon and nitrogen abundances from our \chjwl\ and \cnjwl\ indices. The mean primordial \cfech\ and \nfecn\ abundances are in good agreement in Galactic GCs with comparable ages and slightly metal-poor Galactic GCs, M92 and M30, suggesting that their proto-GC clouds may experienced similar degree of the carbon and nitrogen enrichment.

Our RGBB $V$ magnitude is in good agreement with that of previous study by \citet{testa95}. The RGBB $V$ magnitudes, which sensitively depend on metallicity and helium abundance, of the FG and SG do not show any perceptible difference. It is not clear whether it is due to small number of RGB stars in NGC\,2257 or being less prominence RGBB features with decreasing metallicity \citep{fusipecci90}. However, the same CDRs, which depend on the helium abundance at a given metallicity, between the FG and SG also suggest that the SG of NGC\,2257 is not significantly enhanced in the helium abundance. At the same time, a compact BHB morphology of NGC\,2257 also suggest a very small internal helium abundance dispersion between MPs compared to the second parameter pair, M55.

We investigated the carbon and nitrogen abundance dispersions, \scnfe\ and \scpnfe\ of NGC\,2257 and some Galactic GCs using our photometric system. We found the metallicity gradients against \scnfe\ and \scpnfe\ of the Galactic normal GCs, analogy to the strong correlations between intrinsic RGB pseudo color indices versus metallicity found by \citet{milone17}, while the metal-complex Galactic GCs exhibit deviate significantly from these relations. Our results show that NGC\,2257 is in agreement with the Galactic normal GCs. After removing metallicity effect, we showed that the \dscnfe\ and \dscpnfe\ of the Galactic GCs are correlated with the GC mass \citep[see also][]{milone17}. Similar to the \scnfe\ and \scpnfe\ versus \feh\ relations, NGC\,2257 is in agreement with the Galactic normal GCs.

We compared \dscnfe\ and \dscpnfe\ versus \fgfrac. NGC\,2257 is in agreement with those of Galactic normal GCs.  We found that the fraction of the SG increases with the \dscnfe\ and \dscpnfe, which makes sense because the more massive GCs would experience more active chemical enrichment before the formation of the SG and retain more SG populations.
In this regard, we argued that the degree of the metallicity dependency removed carbon and nitrogen abundance dispersions of the RGB stars fainter than RGBBs, \dscnfe\ and \dscpnfe, and the intrinsic RGB width in pseudo color indices are pseudo--invariant parameters, set at the formation epoch of GC MPs.

We expanded our investigation using the HST observations done by others, finding that the \rgbw, a good proxy of our \dscnfe, is strongly anticorrelated with the \fgfrac\ in the Galactic normal GCs. On the other hand, metal-complex Galactic GCs (and M4) appear to have completely different \rgbw\ versus \fgfrac\ relations than the bulk of Galactic normal GCs, suggesting that the formation of the MPs in these metal-complex GCs and M4 must have been very different from that of normal GCs. We found that NGC\,2257 is in excellent agreement with the relation from the Galactic normal GCs. Finally, we showed that the three old LMC GCs  (NGC\,1786, NGC\,1898, and NGC\,2257) appear to follow the same trends as the Galactic GCs in the \rgbw, \fgfrac, and \logmass\ domains, indicating that the physical environment of the host galaxy did not play a major role in the formation and evolution of GC MPs

It is believed that our ongoing project with our own photometric system with expanded Galactic and extragalactic samples will shed more light in understanding the formation of GC MPs with different physical environments.

\begin{acknowledgements}
We thank the anonymous referee for encouraging and helpful comments.
J.-W.\ Lee acknowledges financial support from the Basic Science Research Program (grant No.\ 2019R1A2C2086290) through the National Research Foundation of Korea (NRF) and from the faculty research fund of Sejong University in 2022.
Based on observations (GS-2022B-Q-105) obtained at the international Gemini Observatory, a program of NSF NOIRLab, which is managed by the Association of Universities for Research in Astronomy (AURA) under a cooperative agreement with the U.S. National Science Foundation on behalf of the Gemini Observatory partnership: the U.S. National Science Foundation (United States), National Research Council (Canada), Agencia Nacional de Investigaci\'{o}n y Desarrollo (Chile), Ministerio de Ciencia, Tecnolog\'{i}a e Innovaci\'{o}n (Argentina), Minist\'{e}rio da Ci\^{e}ncia, Tecnologia, Inova\c{c}\~{o}es e Comunica\c{c}\~{o}es (Brazil), and Korea Astronomy and Space Science Institute (Republic of Korea).
\end{acknowledgements}

\facilities{Gemini-South: 8.0 m (GMOS), KPNO WIYN : 0.9 m (HDI), LCO SWOPE : 1.0 m (CCD)}
\software{\atlas\ \citep{kurucz11}, \moogscat\ \citep{moogscat}, \linemake\ \citep{linemake}}

\end{document}